\documentclass[preprint,showpacs,preprintnumbers,amsmath,amssymb]{revtex4}

\usepackage{epsfig}
\usepackage{graphics}
\usepackage{graphicx}
\usepackage{dcolumn}
\usepackage{bm}

\begin{document}

\preprint{APS/123-QED}

\title{Transition from phase to generalized synchronization in time-delay
systems}

\author{D.~V.~Senthilkumar$^1$}
\email{skumar@cnld.bdu.ac.in}
\author{M.~Lakshmanan$^1$}%
 \email{lakshman@cnld.bdu.ac.in}
 \author{J.~Kurths$^2$}
 \email{jkurths@gmx.de}
\affiliation{%
$^1$Centre for Nonlinear Dynamics,Department of Physics,
Bharathidasan University, Tiruchirapalli - 620 024, India\\
}%
\affiliation{%
$^2$ Institute of Physics, University of Potsdam, 
Am Neuen Palais 10, 14469 Potsdam, Germany\\
}%

\date{\today}

\begin{abstract}  

The notion of phase synchronization in time-delay systems, exhibiting highly
non-phase-coherent attractors, has not been realized yet even though it has
been well studied in chaotic dynamical systems without delay. We report the
identification of phase synchronization in coupled nonidentical piece-wise
linear and in coupled Mackey-Glass time-delay systems with highly
non-phase-coherent  regimes. We show that there is a transition from
non-synchronized behavior to phase and then  to generalized synchronization as
a function of coupling strength. We have introduced a transformation to capture
the phase of the non-phase coherent attractors, which works equally well for
both the time-delay systems.  The instantaneous phases of the above  coupled
systems calculated from the transformed attractors satisfy both the phase and
mean frequency locking conditions. These transitions are also characterized  in
terms of recurrence based indices, namely generalized autocorrelation function
$P(t)$, correlation of probability of recurrence (CPR), joint probability of
recurrence (JPR) and similarity of probability of recurrence (SPR). We have
quantified the different synchronization regimes in terms of these indices. The
existence of phase synchronization is also characterized by  typical
transitions in the Lyapunov exponents of the coupled time-delay systems.

\end{abstract}

\maketitle

{\bfseries Synchronization of chaotic oscillations is one of the most
fundamental phenomena exhibited by coupled chaotic oscillators. Since the
identification of chaotic synchronization in identical systems, several
different kinds of synchronizations such as generalized, phase, lag,
anticipatory and intermittent synchronizations have been identified and
demonstrated.  Among them chaotic phase synchronization (CPS) plays a crucial
role in understanding a large class of weakly interacting nonlinear dynamical
systems. Even though the notion of CPS has been well studied in several
low-dimensional chaotic dynamical systems during the past decade, CPS in
time-delay systems (which are effectively infinite-dimensional) has not yet been
identified and reported.  A main problem here is to define even the notion of
phase itself due to the presence of intrinsic multiple characteristic time
scales of the chaotic attractors. Time-delay systems often exhibit complicated
non-phase-coherent attractors (which do not have proper rotation around a fixed
reference point) with many positive Lyapunov exponents. Hence, the conventional
techniques available in the literature to define phase and to identify CPS
cannot be used  in the case of time-delay systems.  In order to overcome this
difficulty, we have introduced a nonlinear transformation which transforms the
non-phase-coherent chaotic/hyperchaotic attractors of specific time-delay
systems into phase-coherent attractors. The transformed attractors allow for the
use of conventional methods to identify phase and CPS in time-delay systems. We
have also confirmed the onset of phase and the transition from desynchronized
state to phase synchronization and its subsequent transition to generalized
synchronization as a function of coupling strength using recurrence based
indices. These results are also corroborated by the changes in the Lyapunov
exponents of the coupled time-delay systems.}

\section{\label{sec:level1}Introduction}

Synchronization of chaotic oscillations is a fundamental nonlinear phenomenon
observed in diverse areas of science and technology.  Since the first
identification of chaotic synchronization, several types of synchronization
have been identified and demonstrated both theoretically and experimentally
\cite{aspmgr2001,sbjk2002,lp1997,jk2000}.  Complete (or identical) synchronization
\cite{hfty1983,asp1984,lmptlc1990,lmptlc1997}, generalized synchronization
\cite{nfrmms1995,lkup1996} and phase synchronization
\cite{mgrasp1996,tyycl1997,ycl1998} are the three main types of
synchronization that have been characterized by the difference in the degree of
correlation between the interacting chaotic dynamical systems.  Among these,
chaotic phase synchronization (CPS) has become the focus of recent research as
it plays a  crucial role in understanding the behavior of a large class of
weakly interacting dynamical systems in diverse natural systems including
circadian rhythm, cardio-respiratory systems, neural oscillators, population
dynamics, etc \cite{aspmgr2001,sbjk2002,jk2000}. Definition of CPS is a direct
extension of the classical definition of synchronization of periodic
oscillations and can be referred to as entrainment between the phases of
interacting chaotic systems, while the amplitudes remain chaotic and, in
general, non-correlated~\cite{gvoasp1997}~(see also Appendix~\ref{a1}).

The notion of CPS has been investigated so far in oscillators driven by
external periodic force \cite{aspgo1997,aspmgr1997}, chaotic oscillators with
different natural frequencies and/or with parameter mismatches
\cite{mgrasp1996,mgrasp1997,uplj1996,mzgww2002}, arrays of coupled chaotic
oscillators \cite{gvoasp1997,mzzz2000} and  also in essentially different
chaotic systems \cite{ercmt2003,sgchl2005}.  In addition CPS has also been
demonstrated experimentally in various systems, such as electrical
circuits~\cite{ercmt2003,apoc2003,msbtps2003,skdbb2006},
lasers~\cite{kvvvni2001,djdrb2001}, fluids~\cite{dmav2000}, biological
systems~\cite{ptmgr1998,rceals1998}, climatology~\cite{dmjk2005}, etc. On the
other hand CPS in nonlinear time-delay systems, which form an important class
of dynamical systems, have not yet been identified  and addressed. A main
problem here is to define  even the notion of  phase in time-delay systems due
to the intrinsic multiple characteristic time scales in these systems. 
Studying CPS in such chaotic  time-delay systems is  of considerable importance
in many fields,  as  in understanding the behavior of nerve cells
(neuroscience), where memory effects play a  prominent role,  in 
physiological studies, in ecology, in lasers,  etc
~\cite{aspmgr2001,sbjk2002,jk2000,mcmlg1977,thif2003,nkgbe2000,mkph2005,lbsibs2006}.

While studying CPS, one usually encounters with the terminologies
phase-coherent and non-phase-coherent chaotic attractors. If the flow of a
dynamical system has a proper rotation around a fixed reference point, then the
corresponding attractor is termed as phase-coherent attractor.  In contrast, if
the flow does not have a proper rotation around a fixed reference point then the
corresponding attractor is called as non-phase-coherent attractor. (More
discussion on the distinction between the phase-coherent and non-phase-coherent
chaotic attractors along with an illustration is given below in
Appendix~\ref{a1}). While  methods  have been  well established in the
literature to identify phase  and to study CPS in phase-coherent chaotic
attractors (see again Appendix~\ref{a1}), methods to identify phase of
non-phase-coherent chaotic attractors have not yet been well established.  Even
the most promising approach based on the idea of curvature to calculate  the
phase of non-phase-coherent attractors  is limited to low-dimensional systems
and unfortunately methods to identify phase and to study CPS in time-delay
systems which often posses highly complicated hyperchaotic attractors have not
yet been identified and reported.

Recently, we have pointed out briefly the identification of CPS in
unidirectionally coupled nonidentical time-delay systems exhibiting hyperchaos
with highly non-phase-coherent attractors~\cite{dvskml2006}.  In this paper we
present our detailed results on the  identification and existence of CPS in
coupled piecewise-linear time-delay systems and in coupled Mackey-Glass
time-delay systems with parameter mismatches.  We will show the entrainment of
phases of the coupled systems from asynchronous state and its subsequent
transition to generalized synchronization (GS) as a function of coupling
strength.  Phases of these time-delay systems are calculated using the
Poincar\'e method after a newly introduced transformation of the corresponding
attractors, which transforms the original non-phase-coherent attractors of both
the systems into smeared limit cycle like attractors. Further, the existence of
CPS and GS in both  the coupled systems are characterized by  recently proposed
methods based on recurrence quantification analysis and in terms of Lyapunov
exponents of the coupled time-delay systems. Thus, the main results of
our paper are
\begin{enumerate}
\item Suitable nonlinear transformation involving delay time can be  introduced
which transforms a chaotic/hyperchaotic non-phase-coherent attractor to a
phase-coherent attractor.  Then it is easier to find  the onset of CPS, GS, etc.
using these transformed phase-coherent attractors.
\item Recurrence based indices can be directly used to identify phase, CPS, GS
from the original non-phase-coherent chaotic/hyperchaotic attractors.
\item Lyapunov exponents also work as a good guide for the synchronization
transitions involving chaotic /hyperchaotic non-phase-coherent attractors.
\end{enumerate} 

The plan of the paper is as follows. In Sec. II, a brief discussion about the
concept of CPS (the  possibility of estimation of the phase in chaotic
systems is presented in detail in Appendix~\ref{a1}) and details of the
time-delay systems, namely, piece-wise linear time-delay system and
Mackey-Glass system under investigation  are presented.  In Sec. III, we point
out the existence of CPS and GS in unidirectionally coupled  piecewise-linear
time-delay systems using the Poincar\'e section technique (after the introduced
transformation), recurrence quantification analysis and Lyapunov exponents of
the coupled systems.  We will also discuss the
existence of CPS and GS in unidirectionally coupled Mackey-Glass time-delay
systems in Sec. IV, using  the above three different approaches.
Finally in Sec. V, we summarize our results.

\section{CPS and Time-delay systems}

CPS has been studied extensively during the last decade in various nonlinear
dynamical systems as discussed in the introduction.  However, only  a few
methods have been available in the literature ~\cite{aspmgr2001,sbjk2002} (for
more details see Appendix~\ref{a1}) to calculate the phase of chaotic attractors
but unfortunately some of these measures are restricted to phase-coherent
chaotic attractors, while the others to  non-phase-coherent chaotic attractors
of low-dimensional systems. It is to be noted  that these conventional methods
available so far in the literature (as discussed briefly in the
Appendix~\ref{a1}) to identify phase of the phase-coherent/non-phase-coherent
attractors  cannot be used in the case of time-delay systems in general, as such
systems will very often exhibit more complicated attractors with more than one
positive Lyapunov exponents.  Correspondingly methods to calculate the phase of
non-phase-coherent hyperchaotic attractors of time-delay systems are not readily
available.  The most promising approach available in the literature to calculate
the phase of non-phase-coherent attractors is based on the concept of
curvature~\cite{gvobh2003}, but this is often restricted to low-dimensional
systems. However, we find that this procedure does not work in the case of
nonlinear time-delay systems in general, where very often the attractor is
non-phase-coherent and high-dimensional. Hence defining and estimating phase
from the hyperchaotic attractors of the time-delay systems itself is a
challenging task and so specialized techniques/tools have to be identified to
introduce the notion of phase in such systems.

It is to be noted that a variety of other nonlinear techniques such as mutual
information, recurrence analysis, predictability etc. can be used to identify
basic types of synchronization~\cite{nmmcr2007}. In particular, mutual
information, predictability and their variants have been used for
characterizing the existence of complete synchronization, generalized
synchronization and the interdependencies among the measured time series of
dynamical systems~\cite{sbjk2002,rqqja2000,mpvk2001,acgl1991}.
Mutual information can also be used to measure the degree of PS~\cite{czjk2002},
see also Sec.~\ref{ps_trans} below, provided that phase is already defined.  
Recently, recurrence based indices are shown to be excellent
quantifiers~\cite{nmmcr2007} of basic kinds of synchronization including CPS in
low dimensional systems and even in the case of noisy, non-stationary data's. 
However, as far as we know predictability cannot be used either to define or to
identify PS.  In any case these measures have not been used so far to identify
phase or CPS in time-delay systems.

In order to define/estimate phase and CPS in time-delay systems, in this paper
we have introduced three different approaches. Firstly, we have introduced a
nonlinear transformation involving time-delay variable that transforms the
non-phase-coherent attractors  into phase-coherent attractors.  After this
transformation of the original non-phase-coherent attractor, the transformed
attractor allows one to use the conventional techniques. Next, we have used the
recently introduced  recurrence  based indices  for the first time in time-delay
systems to identify the onset of PS and subsequent transition to GS. Finally,
the transition is also confirmed by the changes in the spectrum of Lyapunov
exponents of the coupled time-delay system. Further, we find that all these three
approaches are in good agreement with the indication of onset of CPS. 

As prototypical examples of nonlinear time-delay systems, we consider two
specific models, namely, (i) a piece-wise linear time-delay
system~\cite{hlzh1996,ptkm1998,dvskmlijbc} and (ii) the  Mackey-Glass time-delay
system~\cite{mcmlg1977,jdf1982} and  investigate the existence of CPS in the
corresponding coupled systems.

\subsection{Piece-wise linear time-delay system}
The following scalar first order delay differential equation was
introduced by Lu and He~\cite{hlzh1996} and discussed in detail by Thangavel et
al.~\cite{ptkm1998},
\begin{eqnarray}
\dot{x}(t)&=&-ax(t)+bf(x(t-\tau)),
\label{eq.onea}
\end{eqnarray}
where $a$ and $b$ are parameters, $\tau$ is the time-delay and $f$ is an
odd piecewise linear function defined as
\begin{eqnarray}
f(x)=
\left\{
\begin{array}{cc}
0,&  x \leq -4/3  \\
            -1.5x-2,&  -4/3 < x \leq -0.8 \\
            x,&    -0.8 < x \leq 0.8 \\              
            -1.5x+2,&   0.8 < x \leq 4/3 \\
            0,&  x > 4/3 \\ 
         \end{array} \right.
\label{eqoneb}
\end{eqnarray}

\begin{figure}
\centering
\includegraphics[width=1.05\columnwidth]{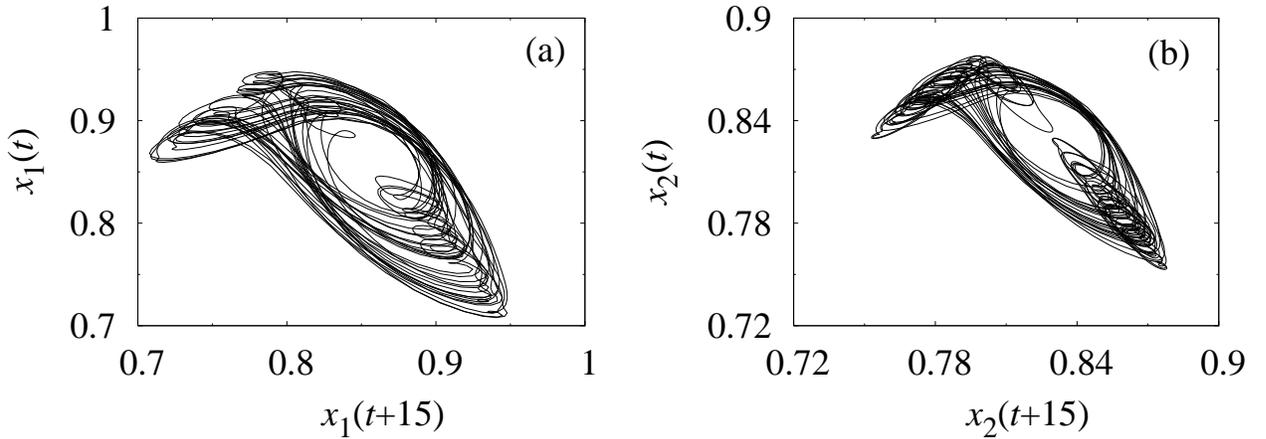}
\caption{\label{fig2b}(a) The non-phase coherent hyperchaotic attractor of the
drive (\ref{eq.one}a) and (b) The non-phase coherent hyperchaotic
attractor of the uncoupled response (\ref{eq.one}b)}
\end{figure}

Recently, we have reported ~\cite{dvskmlijbc}  that systems of the form
~(\ref{eq.onea}) exhibit hyperchaotic behavior for suitable parametric values.
For our present study, we find that for the choice of the parameters $a=1.0,
b=1.2$ and $\tau=15.0$ with the initial condition $x(t)=0.9, t\in(-15,0)$,
Eq.~(\ref{eq.onea}) exhibits hyperchaos.  Detailed linear stability analysis, 
bifurcation analysis and transient effects have been studied in
ref.~\cite{dvskmlijbc}. The corresponding pseudoattractor is shown in the
Fig.~\ref{fig2b}a. The hyperchaotic nature of Eq.~(\ref{eq.onea}) is confirmed
by the existence of multiple positive Lyapunov exponents.   The first ten
maximal Lyapunov exponents for the above choice of parameters as a function of
delay time $\tau \in (2,29)$ are shown in Fig.~\ref{fig1}a (the spectrum of
Lyapunov exponents in this paper are calculated using the procedure suggested
by Farmer \cite{jdf1982}).

Studying synchronization in coupled systems of the form (\ref{eq.onea}) is
particularly appealing because of the facts that (i) system (\ref{eq.onea})
exhibits a hyperchaotic attractor even for very small values of the delay time
$\tau$ for appropriate values of the system parameters (the spectrum of Lyapunov
exponents as a function of delay time $\tau$ is shown in Fig.~\ref{fig1}a) and
(ii) it is easily experimentally realizable as the
piece-wise linear function can be constructed readily and only low values of
delay time are required for construction of a hyperchaotic attractor.

\subsection{Mackey-Glass system}

The second model we have used for the investigation of CPS is a model of blood
production due to Mackey and Glass~\cite{mcmlg1977}. It is represented again by
Eq.~(\ref{eq.onea}) but with the following functional form for $f(x)$

\begin{eqnarray}
f(x)=x(t-\tau)/(1.0+x(t-\tau)^{10}).
\label{eqtwob}
\end{eqnarray}

Here, $x(t)$ represents the concentration of blood at time $t$, when it is
produced, and $x(t-\tau)$ is the concentration when the "request" for more
blood is made.  In patients with leukemia, the time $\tau$ may become
excessively large, and the concentration of blood will oscillate, or if $\tau$
is even larger, the concentration can vary chaotically, as demonstrated by
Mackey and Glass~\cite{mcmlg1977,jdf1982}. This is a prototype model for delay
systems exhibiting highly non-phase-coherent chaotic attractors and even
hyperchaotic attractors for large value of delay time ($\tau >28$). The 
pseudo-chaotic attractor of the Mackey-Glass system (\ref{eq.onea}) and
(\ref{eqtwob}) for the standard parameter values $a=0.1, b=0.2$ and $\tau=20$
with the initial condition $x(t)=0.8, t\in(-20,0)$ is shown in the
Fig.~\ref{fig9}a in Sec.~IV below. The spectrum of Lyapunov exponents as a
function of delay time $\tau \in (14,37)$ is shown in Fig.~\ref{fig8}a (see
Sec.~IV below).

\section{CPS in coupled piecewise-linear time-delay systems}

We first consider the following unidirectionally coupled drive $x_1(t)$ and response
$x_2(t)$ systems, which we have recently studied in detail in
\cite{dvskml2005,dvskml2005jp},
\begin{subequations}
\begin{eqnarray}
\dot{x}_1(t)&=&-ax_1(t)+b_{1}f(x_1(t-\tau)),  \\
\dot{x}_2(t)&=&-ax_2(t)+b_{2}f(x_2(t-\tau))+b_{3}f(x_1(t-\tau)),
\end{eqnarray}
\label{eq.one}
\end{subequations}
where $b_1, b_2$ and $b_3$ are constants, $a>0$, $\tau$ is the delay time and 
$f(x)$ is the piece-wise linear function of the form (\ref{eqoneb}).

\begin{figure}
\centering
\includegraphics[width=1.0\columnwidth]{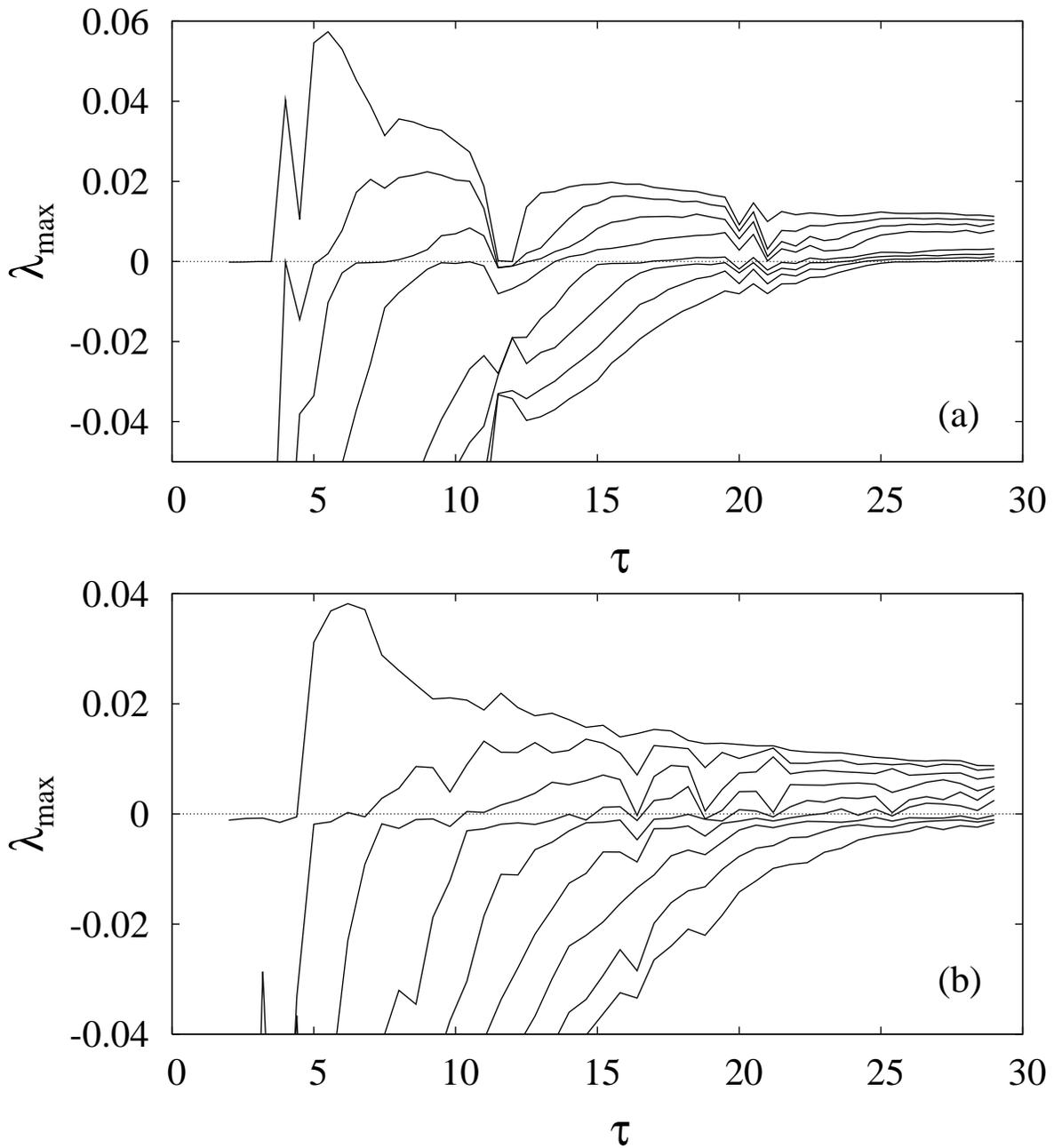}
\caption{\label{fig1} The first ten maximal Lyapunov exponents $\lambda_{max}$
of (a) the scalar time-delay system (\ref{eq.onea}) and (\ref{eqoneb}) or
(\ref{eq.one}a) for the parameter values $a=1.0, b_1=1.2, \tau \in(2,29)$ and
(b) the scalar time-delay system (\ref{eq.one}b) for the parameter values
$a=1.0, b_1=1.1$ in the same range of delay time in the absence of the coupling
$b_3$.}
\end{figure}

We have chosen the values of parameters as (same values as studied in 
ref.~\cite{dvskml2006}) $a=1.0,b_1=1.2,b_2=1.1$ and $\tau=15$. For this
parametric choice, in the absence of coupling,  the drive $x_1(t)$ and the
response $x_2(t)$ systems evolve independently. Further in this case, both the
drive $x_1(t)$  and the response $x_2(t)$ systems exhibit  hyperchaotic 
attractors  with five positive Lyapunov exponents  and  four positive Lyapunov
exponents, respectively, i.e. both subsystems are qualitatively different  (due
to $b_1\ne b_2$).   The corresponding attractors are shown in Figs.~\ref{fig2b}a
and \ref{fig2b}b, respectively, which clearly show the non-phase-coherent
nature. The Kaplan and Yorke~\cite{jdf1982,jkjy1979} dimension for the above
attractors turn out to be  $8.40$ and $7.01$, respectively, obtained by using
the formula
\begin{align}
D_L=j+\frac{\sum_{i=1}^{j}\lambda_i}{\left|\lambda_{j+1}\right|},
\label{lyadim}
\end{align}
where j is the largest integer for which $\lambda_1 + ... + \lambda_j \ge 0$. 
The parameter $b_3$ is the coupling strength of the  unidirectional nonlinear
coupling (\ref{eq.one}b), while the parameters $b_1$ and $b_2$ play the role of
parameter mismatch resulting in nonidentical coupled time-delay systems.  The
spectrum of the first ten largest Lyapunov exponents of the uncoupled system
(\ref{eq.one}a) for the values of the parameters  $a=1.0$ and $b_1=1.2$ in the
range of time-delay $\tau \in (2,29)$ is shown in Fig.~\ref{fig1}a and that of
the system (\ref{eq.one}b) for the parameter value $b_2=1.1$ in the same range
of delay time is also shown in Fig.~\ref{fig1}b.

Now the task is to identify and to characterize the existence of CPS in the
coupled time-delay systems (\ref{eq.one}), possessing highly non-phase-coherent
hyperchaotic attractors, when the coupling is introduced $(b_3>0)$. In the
following we present three different approaches to study CPS in coupled
piecewise-linear time-delay systems (\ref{eq.one}).

\begin{figure}
\centering
\includegraphics[width=1.1\columnwidth]{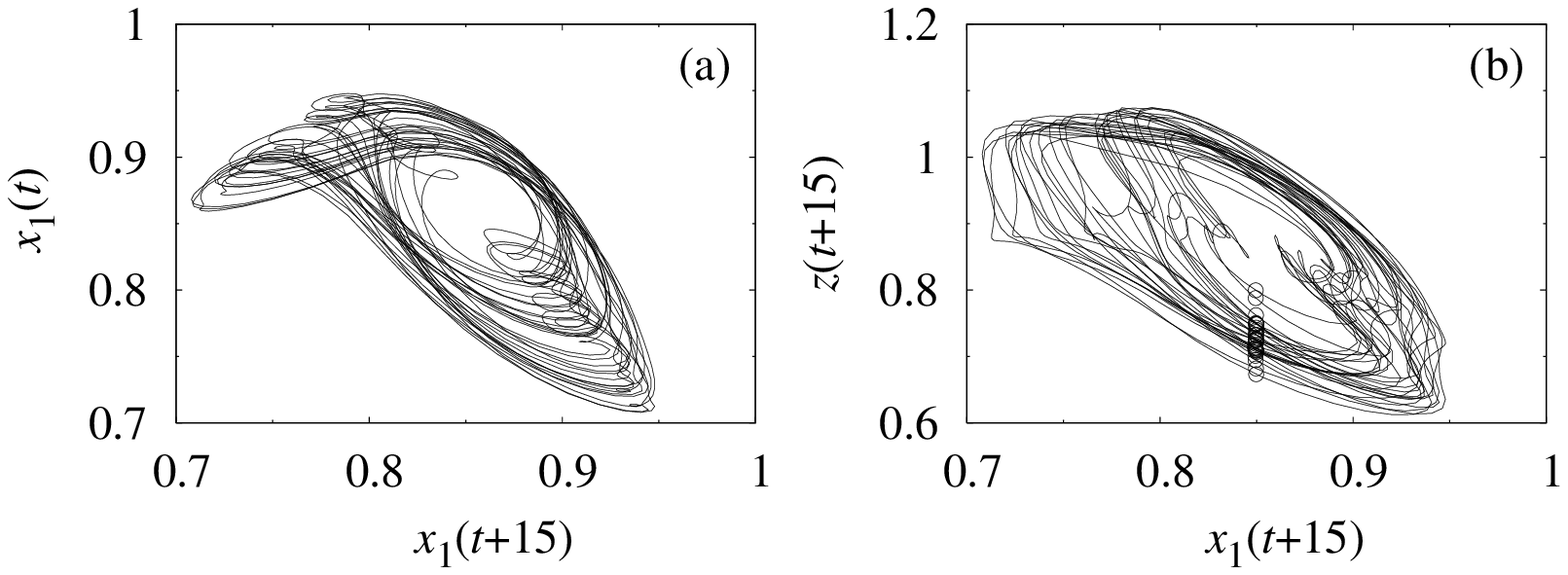}
\caption{\label{chaos}(a) The non-phase coherent hyperchaotic attractor of the
uncoupled drive (\ref{eq.one}a) and (b) Transformed attractor in the
$x_1(t+\tau)$ and $z(t+\tau)$ space. Here  the Poincar\'e points are
represented as open circles.} 
\end{figure}

\subsection{\label{ps_trans}CPS from Poincar\'e section of the transformed attractor
(Fig.~\ref{chaos}b)}

\begin{figure}
\centering
\includegraphics[width=1.0\columnwidth]{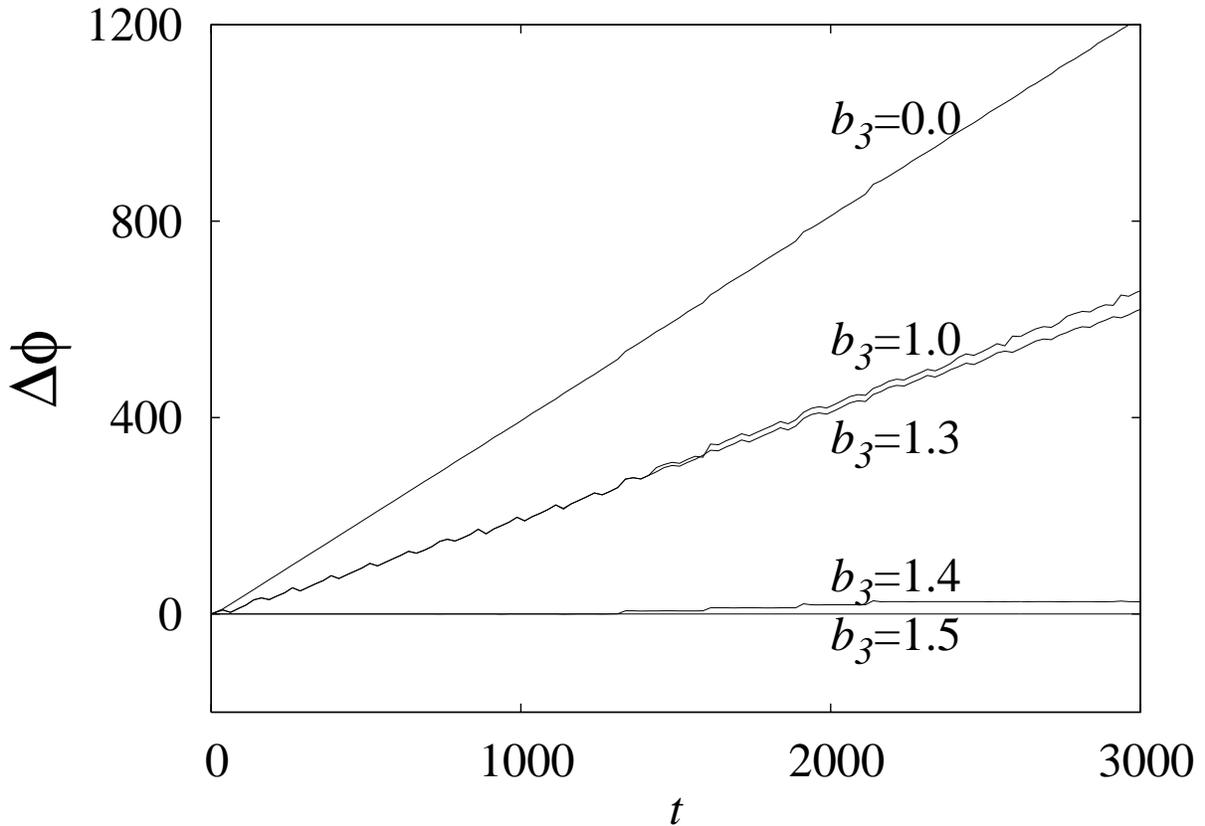}
\caption{\label{fig3} Phase differences ($\Delta \phi =\phi_1^{z}(t)-
\phi_2^{z}(t)$) between the systems (\ref{eq.one}a) and (\ref{eq.one}b) for
different values of the coupling strength $b_3=0.0,1.0,1.3,1.4$ and $1.5$.} 
\end{figure}

We introduce a transformation to successfully capture the phase in the
present problem.    It transforms the non-phase coherent attractor
(Fig.~\ref{chaos}a) into  a smeared limit cycle-like form with well-defined
rotations around one center (Fig.~\ref{chaos}b). This transformation is
performed by introducing the new state variable 
\begin{align}
z(t+\tau)=z(t+\tau,\hat{\tau})=x_1(t)x_1(t+\hat{\tau})/x_1(t+\tau),
\label{trans}
\end{align}
where $\hat{\tau}$ is the optimal value of delay time to be chosen (so as to
rescale the original non-phase coherent attractor into a smeared limit
cycle-like form), and then we plot the above attractor (Fig.~\ref{chaos}a) in
the  ($x_1(t+\tau),z(t+\tau))$ phase space.  The functional form of this
transformation (along with a delay time $\hat{\tau}$) has been identified by
generalizing the transformation used in the case of chaotic atractors in the
Lorenz system \cite{aspmgr2001}, so as to unfold the original non-phase-coherent
attractor  (Fig.~\ref{chaos}a) into a phase-coherent attractor. We find the
optimal value of $\hat{\tau}$ for the attractor (Fig.~\ref{chaos}a) of the
piecewise linear time-delay system to be $1.6$. It is to be noted that on closer
examination of the transformed attractor  (Fig.~\ref{chaos}b) in the vicinity of
the common center, it does not have any closed loop (unlike the case of the
original attractor (Fig.~\ref{chaos}a)) even though the trajectories show sharp
turns in some regime of the phase space. If it is so, such closed loops will
lead to phase mismatch, and one cannot obtain exact matching of phases of both
the drive and response systems as shown in Fig.~\ref{fig3} and discussed below.
Now the attractor (Fig.~\ref{chaos}b) looks indeed like a smeared limit cycle
with nearly well defined rotations around a fixed center. 

It is to be noted that  the above transformation (\ref{trans}) can be applied to
the non-phase-coherent attractors of any  time-delay system in general, except
for the fact that the optimal value of $\hat{\tau}$ should be chosen for each
system appropriately through trial and error by  requiring the geometrical
structure of the transformed attractor to have a fixed center of rotation. We
have adopted here a geometric approach for the selection of  $\hat{\tau}$ and
look for an optimum transform which leads to a  phase-coherent structure.  This
is indeed demonstrated for the attractor of Mackey-Glass system in the next
section. The motivation behind this transformation has came from the
transformation~(\ref{lor}) which is well known in the case of Lorenz attractor
discussed in the Appendix~\ref{a1}. The main point that we want to stress here
is that even for highly non-phase-coherent hyperchaotic attractors of time-delay
systems, there is every possibility to  identify suitable transformations of the
type (\ref{trans}) to unfold the attractor and to identify phase as demonstrated
in the above two typical cases of time-delay systems. One may ask a pertinent
question here as to whether there exists a deeper underlying mathematical
structure regarding such a transform.  We do not have  an answer to this
question at present and this remains an open problem.

Therefore, the phase of the transformed attractor can be now defined based on an
appropriate Poincar\'e section which is transversally crossed by all
trajectories using Eq.~(\ref{phase}) given in Appendix A.  Open circles in 
Fig.~\ref{chaos}b correspond to  the Poincar\'e points of the smeared
limit-cycle-like attractor. Phases, $\phi_1^{z}(t)$ and $\phi_2^{z}(t)$, of  the
drive $x_1(t)$ and the response $x_2(t)$ systems, respectively, are calculated
from the state variables  $z_1(t+\tau)$ and $z_2(t+\tau)$ according to
Eq.~(\ref{trans}). The existence of 1:1 CPS between the systems
(\ref{eq.one}) is characterized by the phase locking condition
\begin{align}
\left|\phi_1^{z}(t)-\phi_2^{z}(t)\right|<const.
\label{ph_con}
\end{align}
The phase differences ($\Delta \phi =\phi_1^{z}(t)- \phi_2^{z}(t)$) between the
systems (\ref{eq.one}a) and (\ref{eq.one}b) are shown in Fig.~\ref{fig3} for
different values of the coupling strength $b_3$. The phase difference $\Delta
\phi$ between the systems (\ref{eq.one}a) and (\ref{eq.one}b) for $b_3=0.0$
(uncoupled) increases monotonically as a function of time confirming that both 
systems are in an asynchronous state (also nonidentical) in the absence of
coupling between them. For the values of $b_3=1.0$ and $1.3$, the phase slips in
the corresponding phase difference $\Delta \phi$ show that the systems are in a
transition state. The strong boundedness of the phase difference specified by
Eq.~(\ref{ph_con}) is obtained for $b_3 > 1.382$ and it becomes zero for the value
of the coupling strength $b_3=1.5$, showing a high quality CPS.   

\begin{figure}
\centering
\includegraphics[width=1.0\columnwidth]{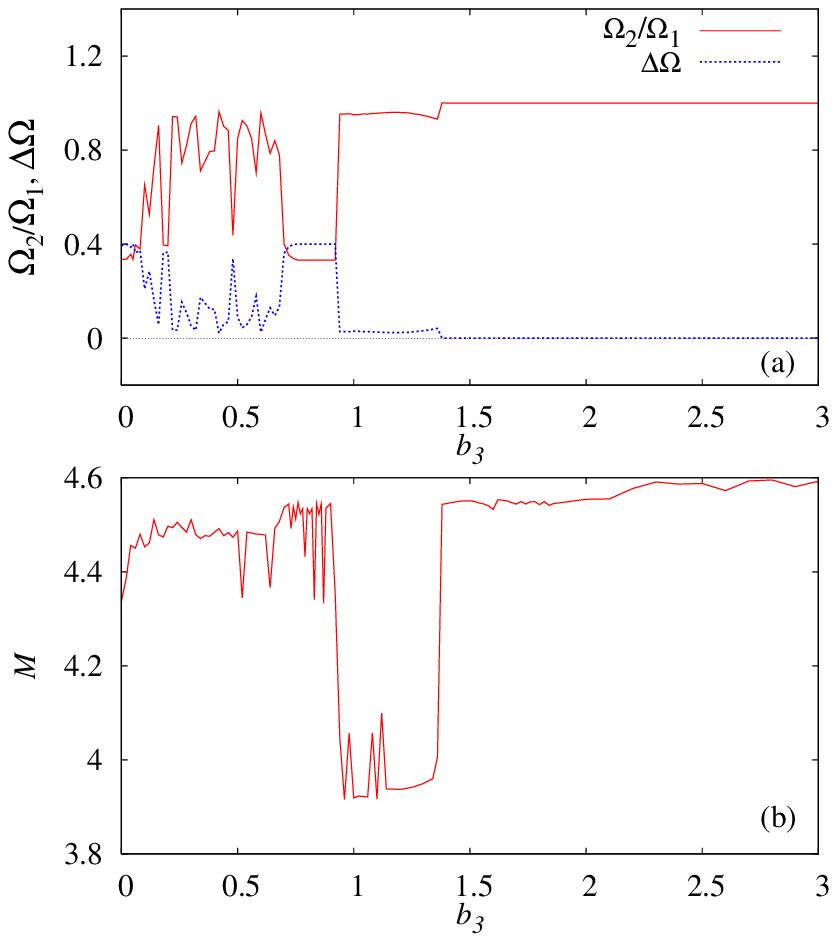}
\caption{\label{fig4}(Color online) (a) Mean frequency ratio $\Omega_2/\Omega_1$
and their difference  $\Delta \Omega=\Omega_2-\Omega_1$ as a function of the
coupling strength $b_3 \in (0,3)$ and (b) Mutual information $M$ as a
function of coupling strength $b_3$.} 
\end{figure}

The mean frequency of the chaotic oscillations is defined as
~\cite{gvoasp1997,sgchl2005}
\begin{align}
\Omega_{1,2}=\langle d\phi_{1,2}^{z}(t)/dt\rangle=\lim_{T\rightarrow\infty}\frac{1}{T}\int_0^T\dot{\phi}_{1,2}(t)dt,
\label{freq}
\end{align}
and the 1:1 CPS between the drive $x_1(t)$ and the response $x_2(t)$  systems
can also be characterized by a weaker condition of frequency locking, that is,
the equality of their mean frequencies $\Omega_1=\Omega_2$.  The mean frequency
ratio $\Omega_2/\Omega_1$ and its difference $\Delta \Omega=\Omega_2-\Omega_1$
are shown in Fig.~\ref{fig4}a as a function of the coupling strength $b_3 \in
(0,3)$.  It is also evident from this figure that the mean frequency locking
criterion (\ref{freq}) is satisfied for $b_3 > 1.382$ from which both the
frequency ratio $\Omega_2/\Omega_1$ and their difference  $\Delta\Omega$  show
substantial saturation in their values confirming the strong boundedness in the
phases of both the systems.

The above results can be further strengthened by  measuring the degree of PS
quantitatively through the concept mutual information between the cyclic
phases~\cite{czjk2002}
\begin{align}
M=\sum_{i,j}p(i,j) ln\frac{p(i,j)}{p_1(i)p_2(j)}, 
\label{mutual}  
\end{align}
where $p_1(i)$ and $p_2(j)$ are the probabilities when the phases $\phi_1$ and
$\phi_2$ are in the $i$th and $j$th bins, respectively, and $p(i,j)$ is the
joint probability that $\phi_1$  is in the $i$th bin and $\phi_2$ in the $j$th
bin. However, it is to be noted that mutual information between the phases
can be used only to characterize the degree of PS provided phase has already
been defined/known.  Hence mutual information can be used only as an additional
quantifier for measuring the degree of phase synchronization. Mutual
information $M$ as a function of coupling strength $b_3 \in (0,3)$ is shown in
Fig.~\ref{fig4}b, which clearly indicates the high degree of PS for $b_3>1.382$
in good agreement with the frequency ratio $\Omega_2/\Omega_1$ and their
difference $\Delta\Omega$ shown in Fig.~\ref{fig4}a.

\subsection{CPS from recurrence quantification analysis}

The complex synchronization phenomena in the coupled time-delay systems 
(\ref{eq.one}) can also be analyzed by means of the very recently proposed
methods based on  recurrence plots \cite{mcrmt2005, nmmcr2007}. These methods help to
identify and quantify CPS  (particularly in non-phase coherent attractors)  and
GS.

For this purpose, the generalized autocorrelation function $P(t)$  has
been introduced in \cite{mcrmt2005, nmmcr2007} as 
\begin{align}
P(t)=\frac{1}{N-t} \sum_{i=1}^{N-t} \Theta(\epsilon-||X_i-X_{i+t}|| ), 
\label{pbt}  
\end{align}
where $\Theta$ is the Heaviside function, $X_i$ is the $i$th data  
corresponding to either the drive variable $x_1$ or the response
variable $x_2$ specified by Eqs.~(\ref{eq.one})
and $\epsilon$ is a predefined threshold. $||.||$ is the  Euclidean
norm and $N$ is the number of data points. $P(t)$ can be considered as a
statistical measure about how often $\phi$ has increased by $2\pi$ or multiples
of $2\pi$ within the time $t$ in the original space.  If two systems are in
CPS, their phases increase on average by $K.2\pi$, where $K$ is a natural
number, within the same time interval $t$. The value of $K$ corresponds to the
number of cycles when $||X(t+T)-X(t)||\sim 0,$ or equivalently when
$||X(t+T)-X(t)|| < \epsilon$, where $T$ is the period of the system. Hence,
looking at the coincidence of the positions of the maxima of $P(t)$ for both 
systems, one can qualitatively identify CPS.

\begin{figure}
\centering
\includegraphics[width=1.1\columnwidth]{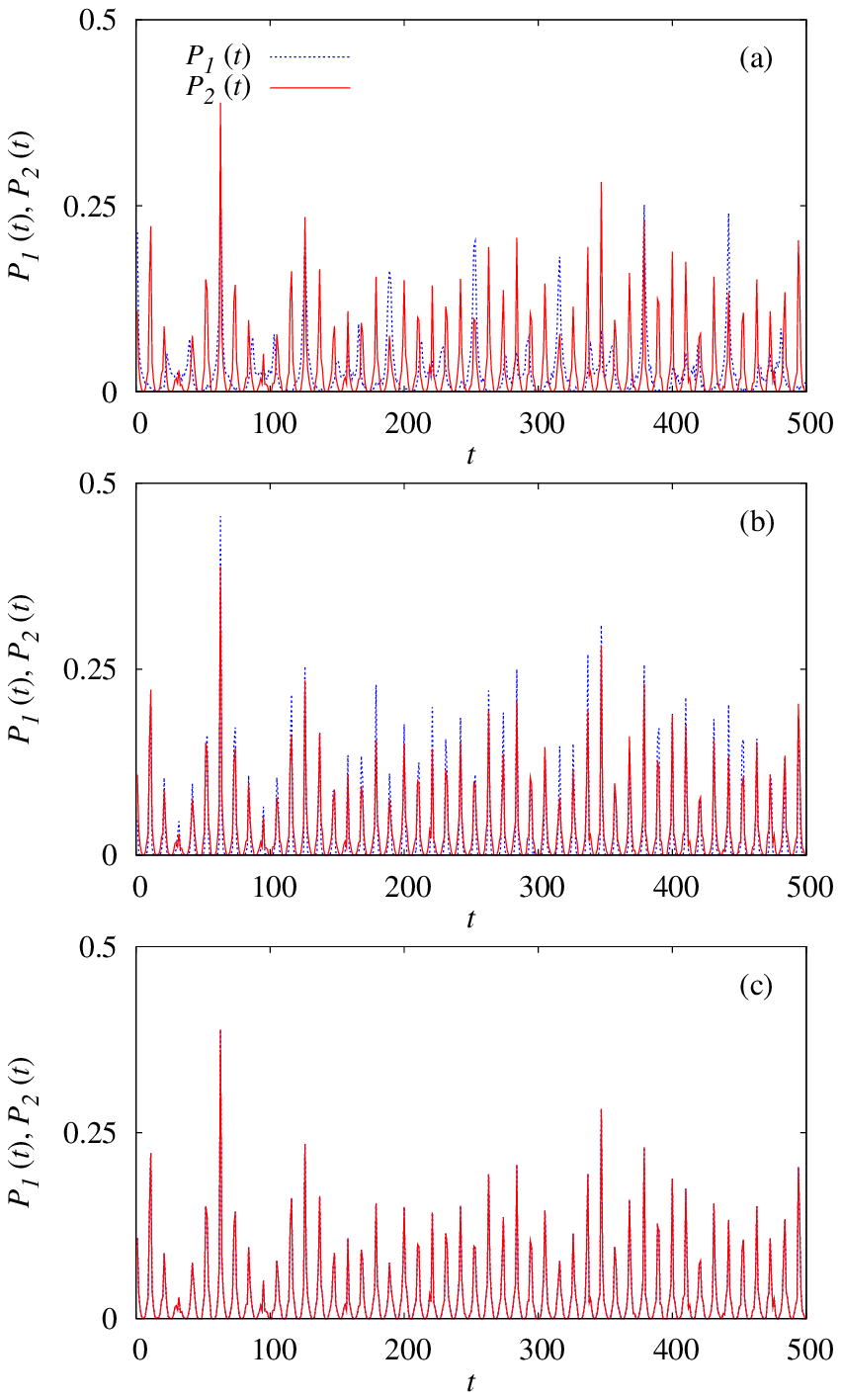}
\caption{\label{fig5}(Color online) Generalized autocorrelation functions of
both the drive $P_1(t)$ and the response $P_2(t)$ systems. (a) Non-phase
synchronization for $b_3=0.6$, (b) Phase synchronization for $b_3=1.5$ and (c)
Generalized synchronization for $b_3=2.3$.}
\end{figure}

A criterion to quantify CPS is the cross correlation 
coefficient between the drive, $P_1(t)$, and the response,
$P_2(t)$, which can be defined as Correlation of Probability of Recurrence
(CPR)
\begin{align}
CPR=\langle \bar{P_1}(t)\bar{P_2}(t)\rangle/\sigma_1\sigma_2,
\end{align}
where $\bar{P}_{1,2}$ means that the mean value has been subtracted and
$\sigma_{1,2}$ are the standard deviations of $P_1(t)$ and $P_2(t)$
respectively.  If both  systems are in CPS, the probability of recurrence is
maximal at the same time $t$ and CPR $\approx 1$. If they are not in CPS,
the maxima do not occur simultaneously and hence one can expect a drift in both
the probability of recurrences and low values of  CPR.

When  the systems (\ref{eq.one}) are in generalized synchronization, two close
states in the phase space of the drive variable correspond to that of the
response.  Hence the neighborhood identity is preserved in phase space.  Since
the recurrence plots are nothing but a record of the neighborhood of each point
in the phase space, one can expect that their respective recurrence plots are
almost identical.  Based on these facts two indices are defined to quantify GS.

First, the authors of \cite{mcrmt2005} proposed the 
 Joint Probability of Recurrences (JPR),
\begin{align}
JPR=\frac{\frac{1}{N^2} \sum_{i,j}^N \Theta(\epsilon_x-||X_i-X_j||
)\Theta(\epsilon_y-||Y_i-Y_j||)-RR}{1-RR}  
\label{jpr}  
\end{align}
where $RR$ is rate of recurrence, $\epsilon_x$ and $\epsilon_y$ are thresholds
corresponding to the drive and response systems respectively and $X_i$ is the $i$th
data corresponding to the drive variable $x_1$ and $Y_i$ is the $i$th data
corresponding to the response variable $x_2$ specified by Eqs.~(\ref{eq.one}). RR
measures the density of recurrence points and it is fixed as 0.02
\cite{mcrmt2005}. JPR is close to $1$ for systems in GS and is small when they
are not  in GS. The second index depends on the coincidence of the  probability
of recurrence, which is defined as Similarity of Probability of Recurrence
(SPR),
\begin{align}
SPR=1-\langle(\bar{P_1}(t)-\bar{P_2}(t))^2\rangle/\sigma_1\sigma_2.
\end{align}  
SPR is of order $1$ if both systems are in GS and approximately zero or
negative if they evolve independently.

\begin{figure}
\centering
\includegraphics[width=1.1\columnwidth]{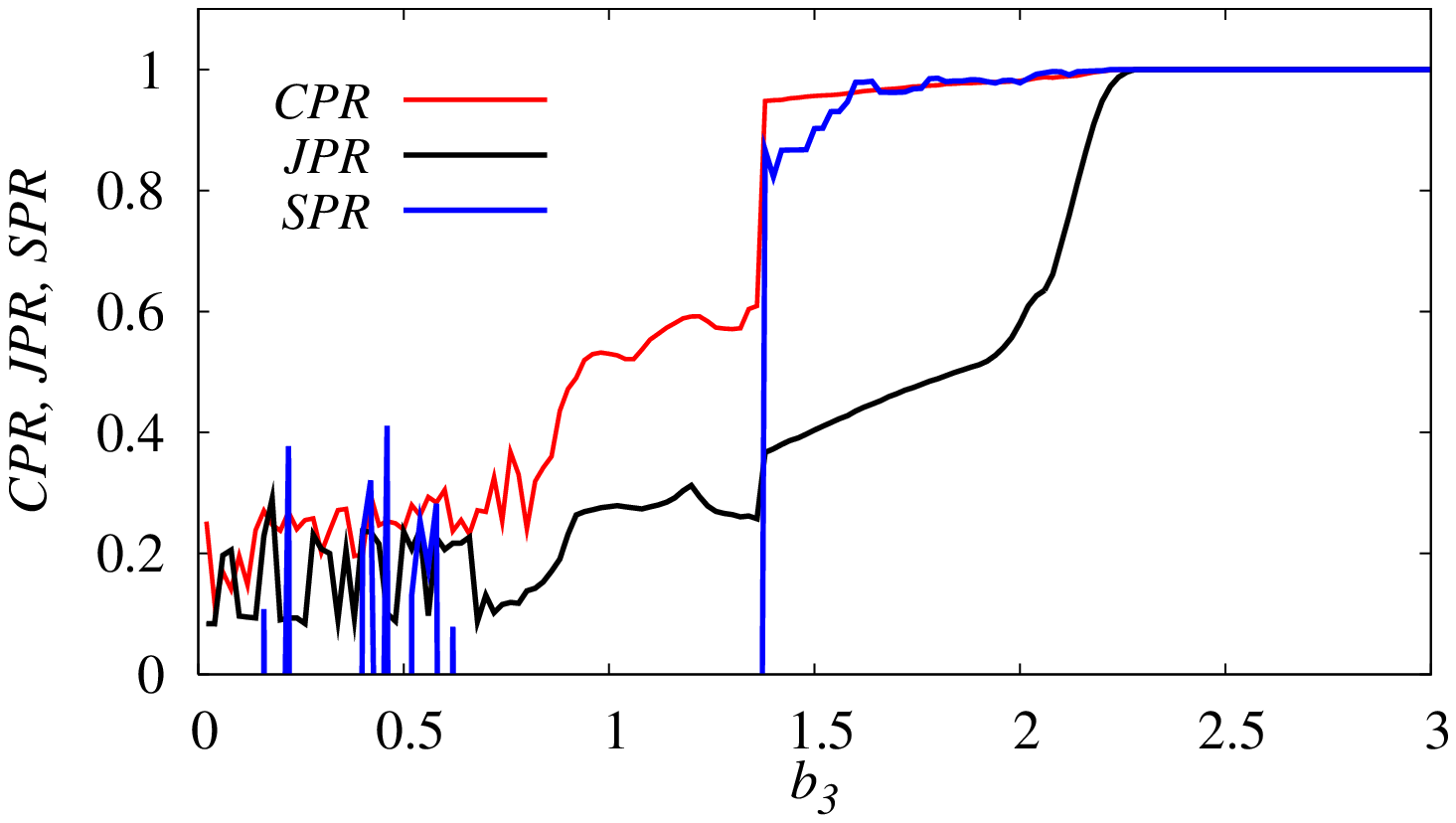}
\caption{\label{fig6}(color online) Indices CPR, JPR and SPR as a function
of coupling strength $b_3 \in (0,3)$.}
\end{figure}

Now, we will apply these concepts to the original (non-transformed) attractor
(Fig.~\ref{chaos}a). We  estimate these recurrence based measures from $5000$
data points after sufficient transients with the integration step $h=0.01$ and
sampling rate $\Delta t =100$. The generalized autocorrelation functions
$P_1(t)$ and $P_2(t)$ (Fig.~\ref{fig5}a) for the coupling $b_3=0.6$ show that
the maxima of both systems do not occur simultaneously and there exists a drift
between them, so there is no synchronization at all.  This is also reflected in
the rather low value of CPR $=0.381$.  For $b_3 \in (0.78,1.381)$, from
Fig.~\ref{fig6} we observe the first substantial increase of recurrence reaching
CPR $\approx 0.5-0.6$. Looking into the details of  the generalized correlation
functions  $P(t)$, we find that now the main oscillatory dynamics becomes
locked, i.e. the main maxima of $P_1$ and $P_2$ coincide. For $b_3 \in
(1.382,2.2)$ CPR reaches almost $1$ as seen in Fig.~\ref{fig6}, while now the
positions of all maxima of $P_1$ and $P_2$ are also in agreement and this is in
accordance with strongly bounded nature of phase differences.  This is a strong
indication for CPS.  Note, however that the heights of the peaks are clearly
different (Fig.~\ref{fig5}b).  The differences in the peak heights indicate that
there is no strong interrelation in the amplitudes. Further increase  of the
coupling (here $b_3=2.21$) leads to the coincidence of both the positions and
the heights of the peaks (Fig.~\ref{fig5}c) referring to GS in systems
(\ref{eq.one}). This is also confirmed from the maximal values of the indices
JPR $=1$ and SPR $=1$, which is due to the strong correlation in the amplitudes
of both systems.  It is clear from the construction of SPR that it measures the
similarity between the generalized autocorrelation functions $P_1(t)$ and
$P_2(t)$. In the regimes of CPS, as the generalized autocorrelation functions
coincide in almost all the regimes except for the height of its  maxima,  it is 
also quantified by  larger values of SPR. The index SPR in Fig.~\ref{fig6} also
shows the onset of CPS and it fluctuates around the value 1 in the regime of CPS
($b_3 \in(1.382,2.2))$  before reaching saturation confirming the strong
correlation in the amplitudes of both the systems, thereby quantifying the
existence of GS. The transition  from non-synchronized state via CPS  to GS is
characterized by the maximal values of  CPR, SPR and JPR (Fig.~\ref{fig6}). As
expected from the construction of these functions, CPR refers mainly to the
onset of CPS, whereas JPR quantifies clearly the onset of GS and SPR indicates
both the onset of CPS and GS. In this connection, we have also confirmed the
onset and existence of GS  by using the auxiliary system
approach~\cite{hdianfr1996} introduced by Abarbanel et al for the range of the
coupling strength $b_3 > 2.2$.

\begin{figure} 
\centering 
\includegraphics[width=1.1\columnwidth]{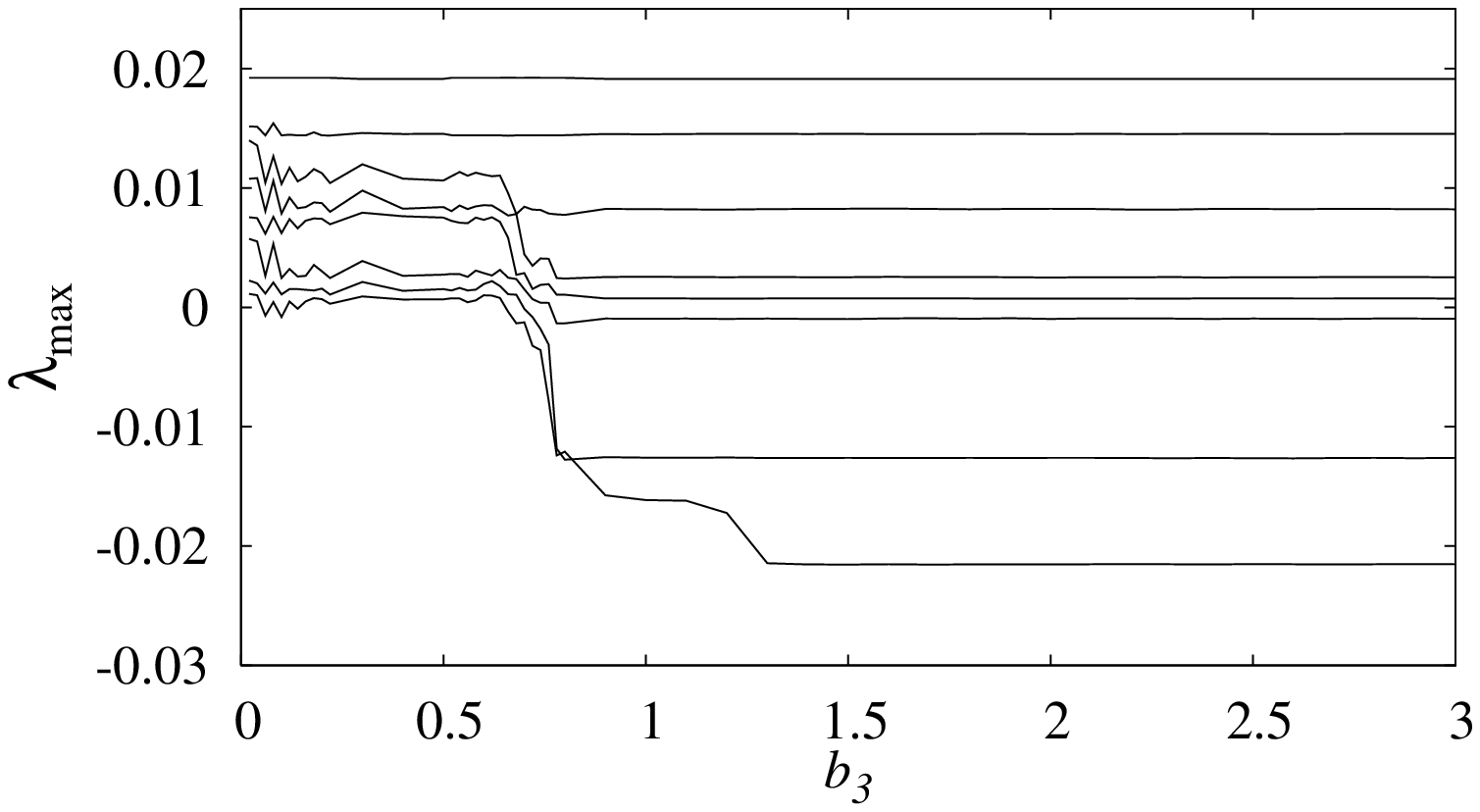}
\caption{\label{fig7} Spectrum of first eight largest Lyapunov exponents  of the
coupled  systems (\ref{eq.one}) as a function of coupling strength $b_3 \in
(0,3)$.} 
\end{figure} 

\subsection{CPS from spectrum of Lyapunov exponents}

The transition from non-synchronization to CPS is also characterized by changes
in the Lyapunov exponents of the coupled time-delay systems (\ref{eq.one}). The
spectrum of the  eight largest Lyapunov exponents of the coupled systems is
shown in Fig.~\ref{fig7}. From this figure one can find that all the positive
Lyapunov exponents, except the largest one ($\lambda_{max}^{(2)}$), 
corresponding to the response system suddenly  become negative  at the value of
the coupling strength $b_3=0.78$ which is an indication of the onset of
transition regime. One may also note that at this value of $b_3$ already one of
the Lyapunov exponents of the response system attains negative saturation while
the another one reaches negative saturation slightly above $b_3=0.78$. This is a
strong indication that in this rather complex attractor the amplitudes become
somewhat interrelated already at the transition to CPS (as in the funnel
attractor \cite{gvobh2003} of the R\"ossler system).  Also the third positive
Lyapunov exponent of the response system gradually becomes more negative from
$b_3=0.78$ and reaches its saturation value at $b_3=1.381$ confirming the onset
of CPS (which is also indicated by the transition of the indices of CPR and SPR
in Fig.~\ref{fig6} in the range of $b_3 \in (0.78,1.381)$).  It is interesting
to note that the Lyapunov exponents of the response system $\lambda_i^{(2)}$
(other than $\lambda_{max}^{(2)}$) are changing already at the early stage of
CPS ($b_3 \in (0.78,1.381)$), where  the  complete CPS  is not yet attained. 
This  has also been observed for the onset of CPS in phase-coherent and
non-phase-coherent oscillators without time-delay~\cite{mgrasp1996,sgchl2005,bhgvo2003}.

\section{CPS in coupled Mackey-Glass systems}
In this section, we will bring out the existence of CPS in coupled Mackey-Glass
systems of the form
\begin{subequations}
\begin{align}
\dot{x}_1(t)=&\,-ax_1(t)+b_{1}x_1(t-\tau)/(1.0+x_1(t-\tau)^{10}),  \\
\dot{x}_2(t)=&\,-ax_2(t)+b_{2}x_2(t-\tau)/(1.0+x_2(t-\tau)^{10})\nonumber \\
&\,+b_{3}x_1(t-\tau)/(1.0+x_1(t-\tau)^{10}),
\end{align}
\label{eqtwo}
\end{subequations}
where $a, b_1, b_2$ are constants, $b_3$ is the coupling parameter, and $\tau$
is the delay time.

\begin{figure}
\centering
\includegraphics[width=1.0\columnwidth]{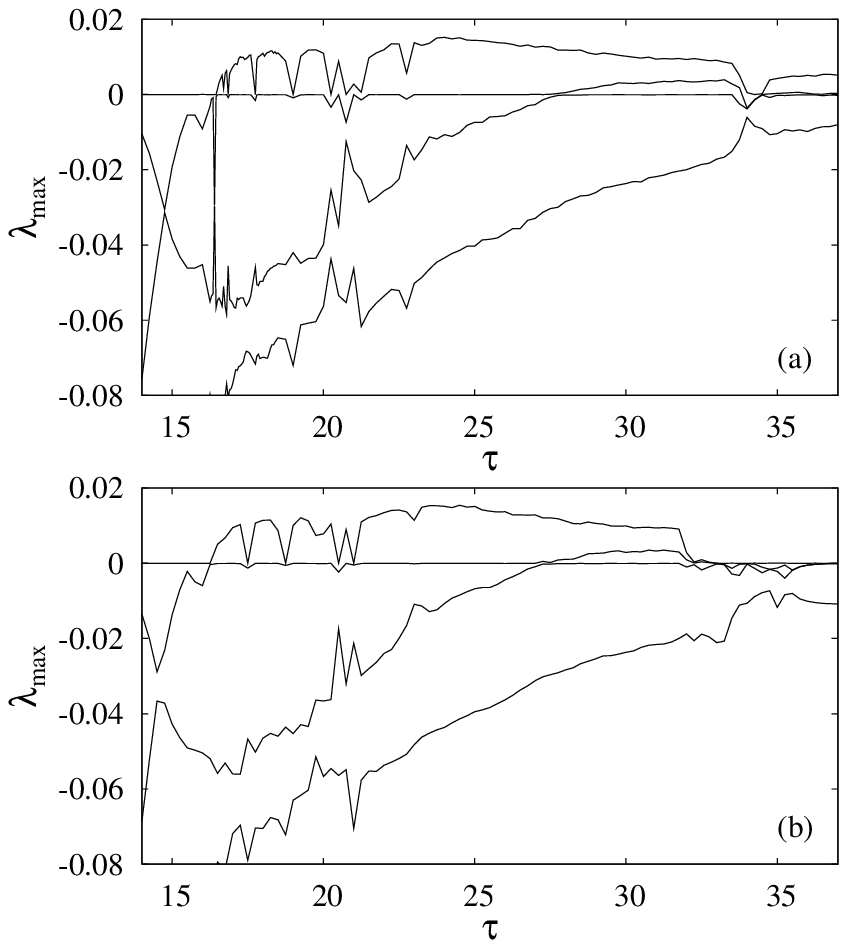}
\caption{\label{fig8} The first four maximal Lyapunov exponents $\lambda_{max}$
of (a) the Mackey-Glass time-delay system (\ref{eqtwo}a) for the parameter
values $a=0.1, b_1=0.2, \tau \in(14,37)$ and (b) time-delay system
(\ref{eqtwo}b) for the parameter values $a=0.1, b_1=0.205$ in the same range of
delay time in the absence of the coupling $b_3$.}
\end{figure}

\begin{figure}
\centering
\includegraphics[width=1.05\columnwidth]{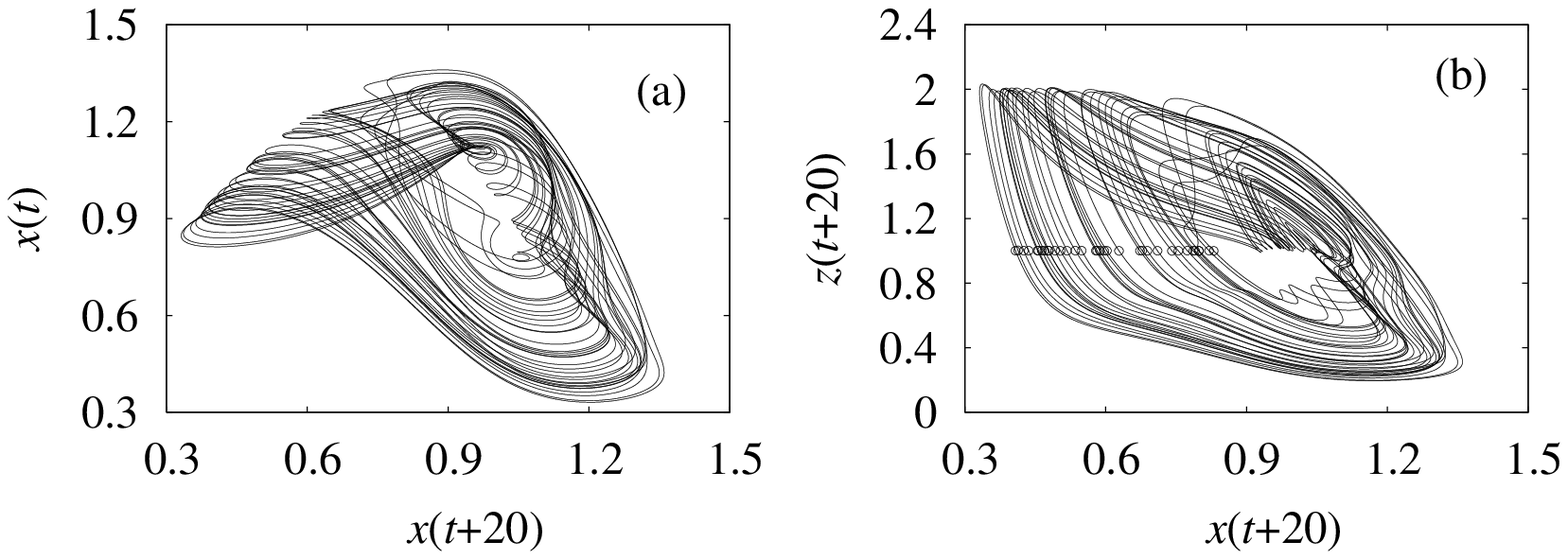}
\caption{\label{fig9}(a) The non-phase coherent chaotic attractor of the
uncoupled drive (\ref{eqtwo}a) and (b) Transformed attractor in the
$x_1(t+\tau)$ and $z(t+\tau)$ space along  with the Poincar\'e points
represented as open circles.}
\end{figure}

We have chosen the parameter values (cf.~\cite{mcmlg1977,jdf1982}) as $a=0.1,
b_1=0.2, b_2=0.205$, $\tau=20$ and varied  the coupling strength $b_3$. The
non-phase-coherent chaotic attractor of the system $x_1(t)$, Eq.~(\ref{eqtwo}a),
for the above choice of parameters is shown in Fig.~\ref{fig9}a and it possesses
one positive and one zero Lyapunov exponents. Similarly, the second system
$x_2(t)$, Eq.~(\ref{eqtwo}b), also exhibits a non-phase-coherent chaotic attractor
with one positive and one zero Lyapunov exponents for the chosen parametric
values in the absence of the coupling strength $b_3$. The parameters $b_1$ and
$b_2$ contribute to the parameter mismatch between the systems $x_1(t)$ and
$x_2(t)$. The spectrum of the first four maximal Lyapunov exponents of both 
systems (\ref{eqtwo}a) and (\ref{eqtwo}b) are shown in Figs.~\ref{fig8}a and
~\ref{fig8}b respectively as a function of delay time $\tau \in (14,37)$ when
$b_3=0$. The Kaplan and Yorke~\cite{jdf1982,jkjy1979} dimension calculated using
(\ref{lyadim}) for  the present systems ((\ref{eqtwo}a) and (\ref{eqtwo}b)) work
out to be $2.27969$ and $2.21096$, respectively.  Now, the
existence of CPS as a function of the coupling strength in the coupled
Mackey-Glass systems (\ref{eqtwo}) will be discussed using the above three
approaches used for identifying CPS in coupled piecewise-linear time-delay
systems (\ref{eq.one}).

\subsection{CPS from Poincar\'e section of the transformed attractor
(Fig.~\ref{fig9}b)}

\begin{figure}
\centering
\includegraphics[width=1.0\columnwidth]{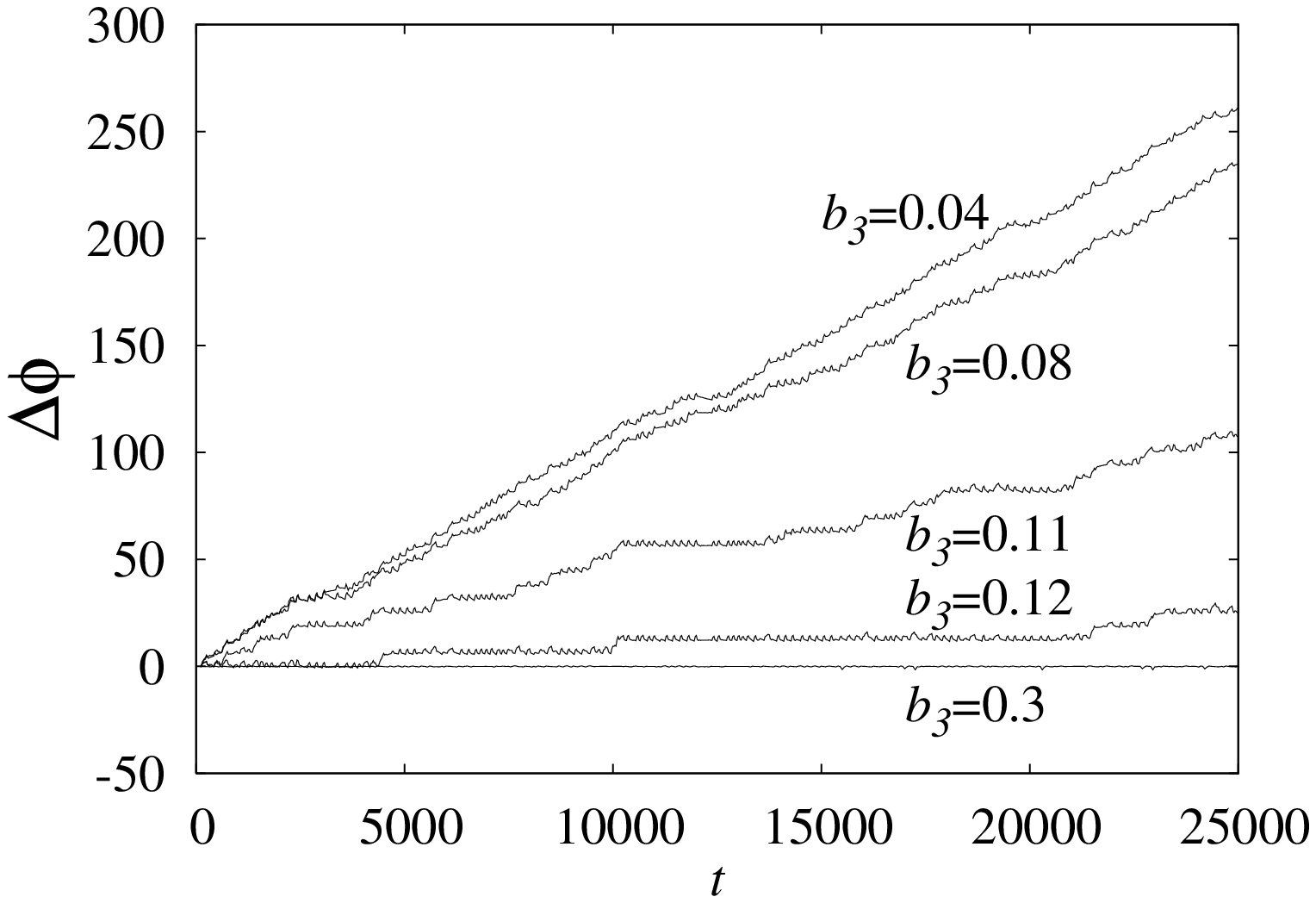}
\caption{\label{fig10} Phase differences ($\Delta \phi =\phi_1^{z}(t)-
\phi_2^{z}(t)$) between the systems (\ref{eqtwo}a) and (\ref{eqtwo}b) for
different values of the coupling strength $b_3=0.04,0.08,0.11,0.12$ and $0.3$.} 
\end{figure}

\begin{figure}
\centering
\includegraphics[width=1.05\columnwidth]{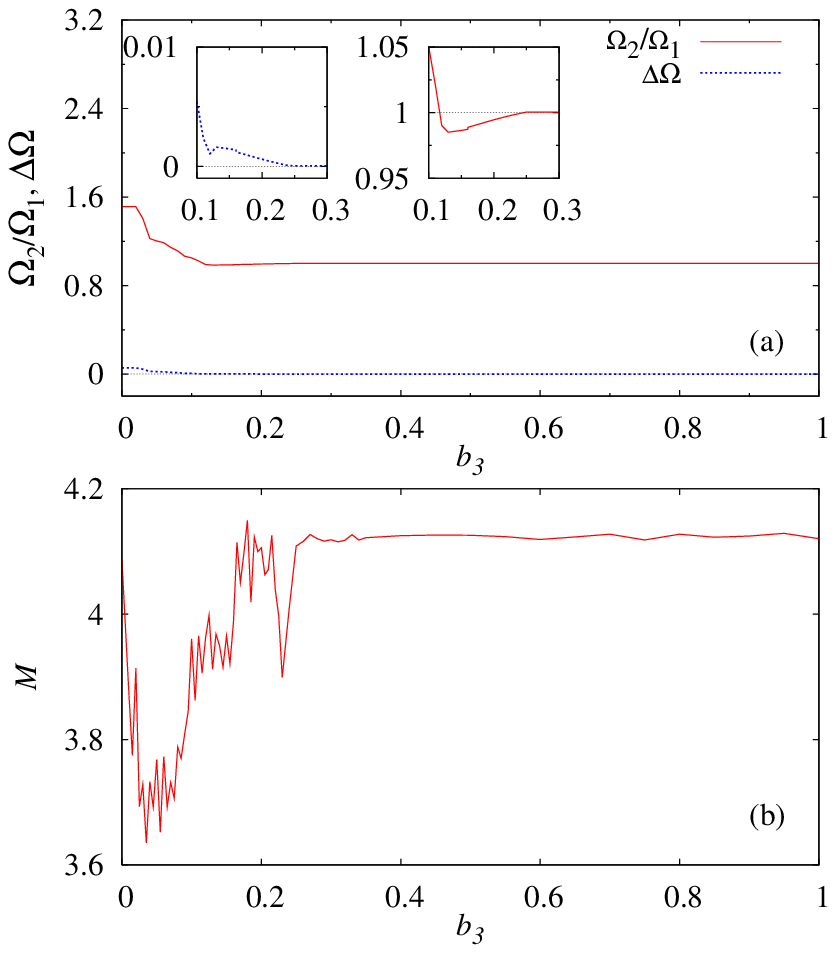}
\caption{\label{fig11}(Color online) (a) Mean frequency ratio $\Omega_2/\Omega_1$
and their difference  $\Delta \Omega=\Omega_2-\Omega_1$ as a function of the
coupling strength $b_3 \in (0,1)$ and (b) Mutual information $M$ as a
function of coupling strength $b_3$.} 
\end{figure}

The non-phase-coherent chaotic attractor (Fig.~\ref{fig9}a) of the Mackey-Glass
system is transformed into a smeared limit cycle-like attractor
(Fig.~\ref{fig9}b) using the same transformation (\ref{trans}) as used for the
piecewise-linear time-delay systems. For the attractor (Fig.~\ref{fig9}a) of the
Mackey-Glass system, the optimal value of the delay time $\hat{\tau}$ in
Eq.~(\ref{trans}) is found to be 8.0. The Poincar\'e points are shown as open
circles in the Fig.~\ref{fig9}b from which the instantaneous phase $\phi_1^z(t)$
is calculated using (\ref{phase}). The existence of CPS in the coupled
Mackey-Glass systems (\ref{eqtwo}) is also characterized by the phase locking
condition (\ref{ph_con}) as shown in Fig.~\ref{fig10}. The phase differences
$\Delta \phi =\phi_1^{z}(t)- \phi_2^{z}(t)$ between the systems (\ref{eqtwo}a)
and (\ref{eqtwo}b) for the values of the coupling strength
$b_3=0.04,0.08,0.11,0.12$ and $0.3$ are shown in Fig.~\ref{fig10}. For the value
of the coupling strength $b_3=0.3$, there exists a strong boundedness in the
phase difference showing high quality CPS. The mean frequency ratio
$\Omega_2/\Omega_1$  calculated from (\ref{freq}) along with the mean frequency
difference $\Delta \Omega$ is shown in the Fig.~\ref{fig11}a. The value of mean
frequency ratio $\Omega_2/\Omega_1 \approx 1$ in the range of $b_3 \in
(0.12,0.23)$ corresponding to the transition regime (which is also to be
confirmed from the indices CPR and JPR in the next subsection), see the inset of
Fig.~\ref{fig11}a.  Similarly the mean frequency difference is also $\Delta
\Omega \approx 0$ confirming the transition regime. For the value of $b_3>0.23$
both quantities $\Omega_2/\Omega_1$ and  $\Delta \Omega$ acquire the complete
saturation in their values confirming the existence of CPS. Further, the mutual
information calculated using Eq.~(\ref{mutual}) clearly indicates the increase
in the degree of PS for the value of coupling strength $b_3>0.23$ as shown in
Fig.~\ref{fig11}b, which is also in agreement with the frequency ratio
$\Omega_2/\Omega_1$ and the mean frequency difference $\Delta \Omega$ shown in
Fig.~\ref{fig11}b.

\subsection{CPS from recurrence quantification analysis}

\begin{figure}
\centering
\includegraphics[width=1.1\columnwidth]{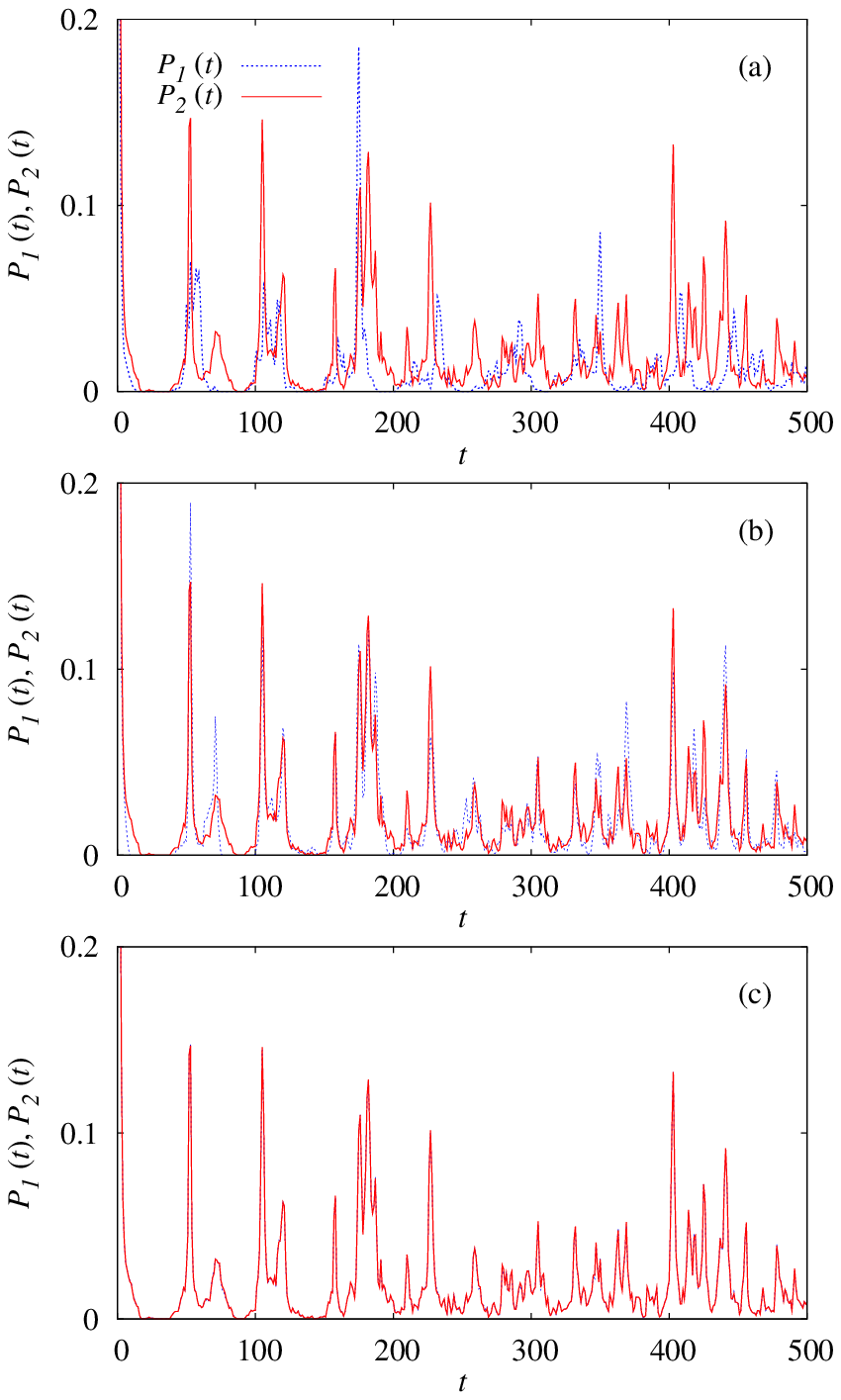}
\caption{\label{fig12}(Color online) Generalized autocorrelation functions of
both the drive system (\ref{eqtwo}a), $P_1(t)$, and the response system
(\ref{eqtwo}b), $P_2(t)$. (a) Non-phase synchronization for $b_3=0.1$,
(b) Phase synchronization for $b_3=0.3$ and (c) Generalized synchronization for
$b_3=0.9$.}
\end{figure}

\begin{figure}
\centering
\includegraphics[width=1.1\columnwidth]{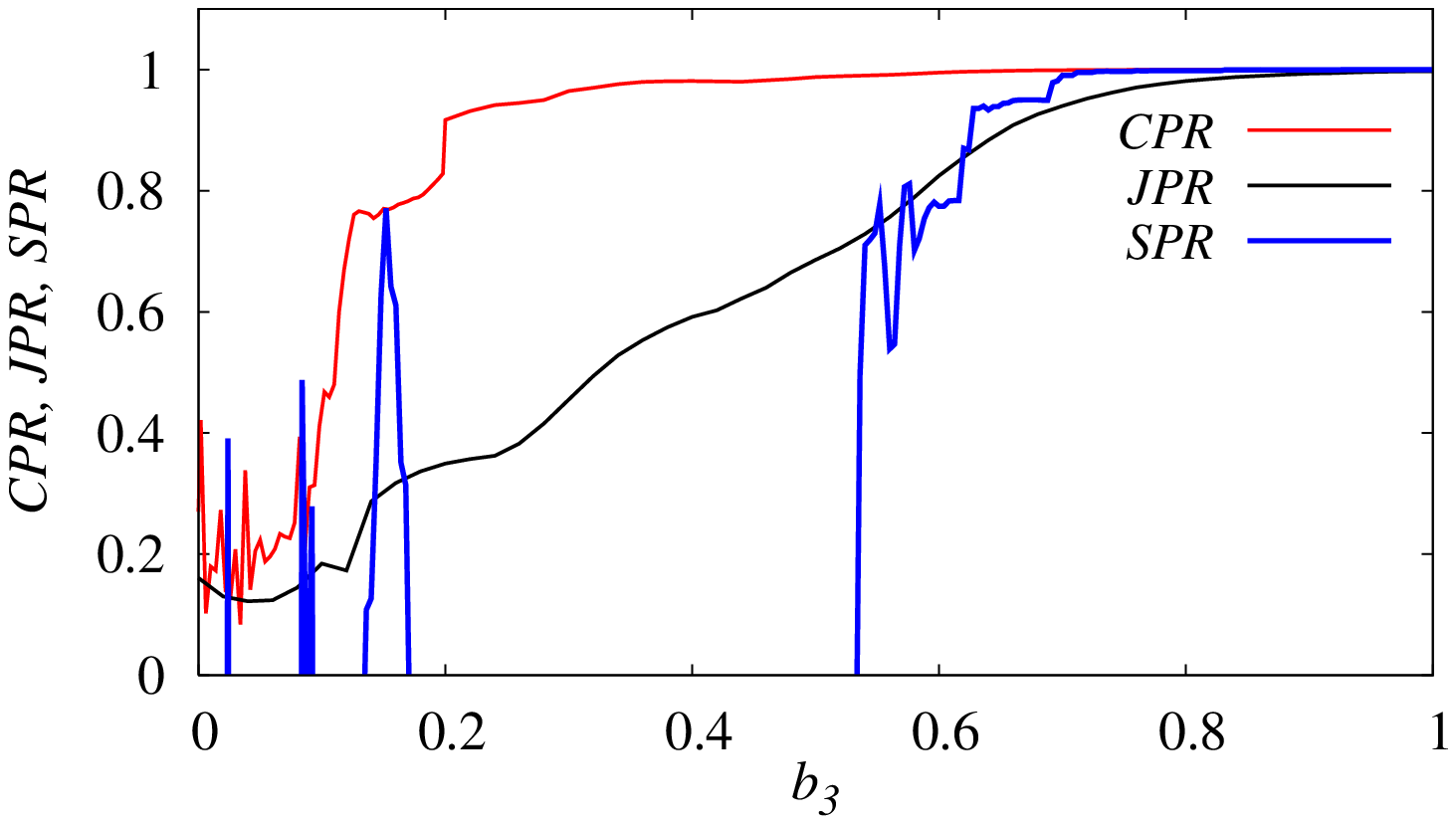}
\caption{\label{fig13} (Color online) Indices CPR, JPR and SPR as a
function of coupling strength $b_3 \in (0,1)$.}
\end{figure}

The existence of CPS from the original non-phase-coherent chaotic attractors of
the systems (\ref{eqtwo}) is analyzed in this section using the recurrence
quantification measures defined in the section II.~B.  We have estimated these
measures again using a set of $5000$ data points, and the same integration step
and the sampling rate as used in the case of coupled piecewise-linear time-delay
systems (\ref{eq.one}). The maxima of  generalized autocorrelations of both the
drive $P_1(t)$ and the response $P_2(t)$ systems (Fig.~\ref{fig12}a) do not
occur simultaneously for $b_3=0.1$, which indicates the independent evolution of
both the systems without any correlation and this is also reflected in the
rather low value of CPR = 0.4. For $b_3=0.3$, the maxima of both $P_1(t)$ and
$P_2(t)$ are in good agreement (Fig.~\ref{fig12}b) and this shows the strongly
bounded phase difference.  It is to be noted that even though both  the maxima
coincide,  the heights of the peaks are clearly of different magnitudes
contributing to the fact that there is no strong correlation in the amplitudes
of both the systems indicating CPS. Both the positions and the peaks are in
coincidence (Fig.~\ref{fig12}c) for the value of coupling strength $b_3=0.9$ in
accordance with the strong correlation in the amplitudes of both the systems
(\ref{eqtwo}) corresponding to GS.  This is also reflected in the maximal values
of both  JPR=1 and SPR=1. The spectra of CPR, JPR and SPR are shown in
Fig.~\ref{fig13}.  The onset of CPS is shown by the first substantial increase
of the index CPR at $b_3=0.11$ and the transition regime is shown by the
successive plateaus of CPR in the range $b_3 \in (0.12,0.23)$.  The maximal
values of CPR  for $b_3>0.23$ indeed confirm the existence of high quality CPS.  The
existence of GS is also confirmed  from both the indices JPR and SPR.  

\subsection{CPS from spectrum of Lyapunov exponents}

\begin{figure}
\centering
\includegraphics[width=1.1\columnwidth]{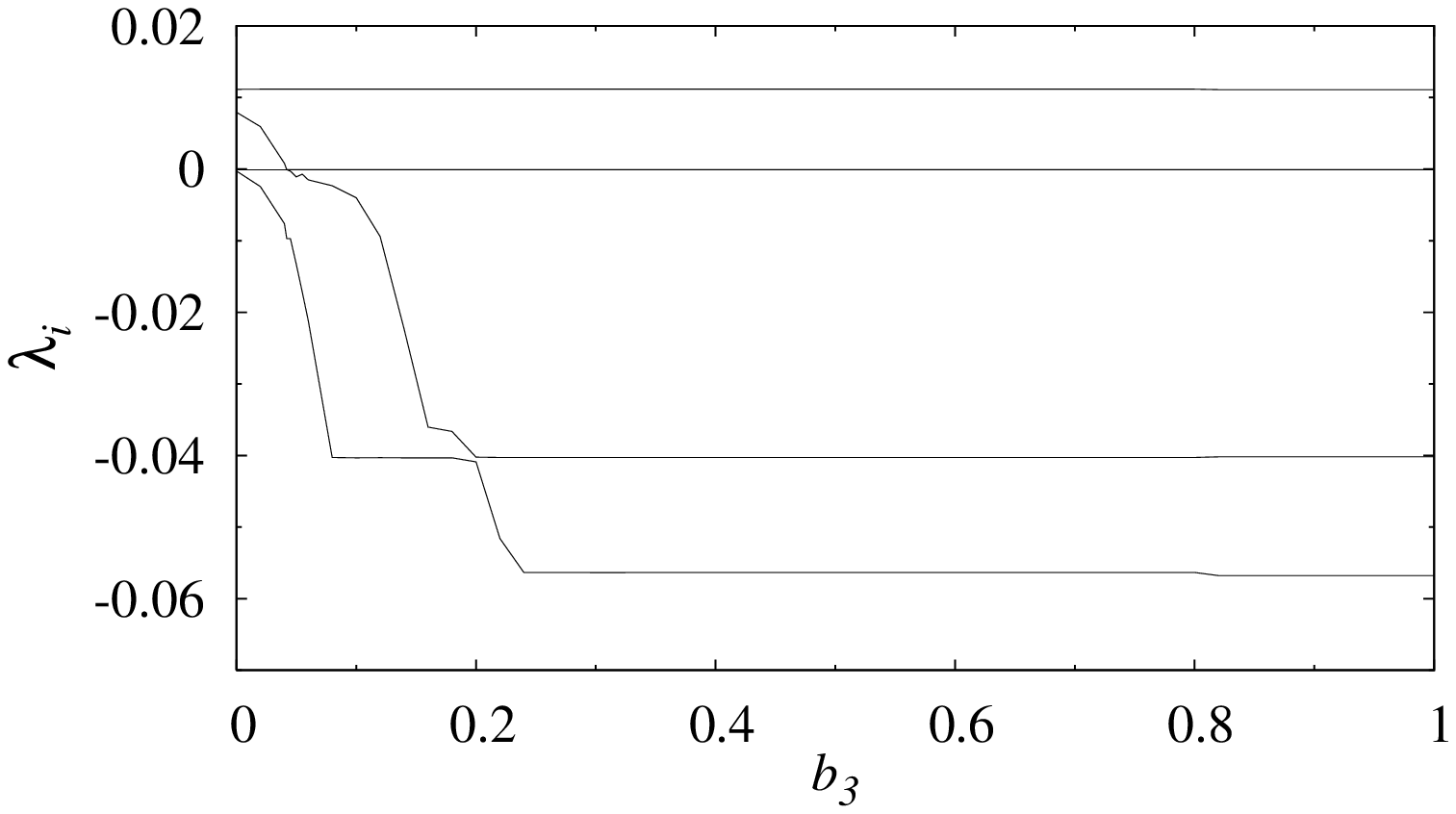}
\caption{\label{fig14}  Spectrum of first four largest
Lyapunov exponents  of the coupled  systems (\ref{eqtwo}) as a function of
coupling strength $b_3 \in (0,1)$.}
\end{figure}

The onset of CPS is also characterized by the changes in the spectrum of
Lyapunov exponents of the coupled Mackey-Glass systems (\ref{eqtwo}). The
spectrum of the first four largest Lyapunov exponents of the coupled systems
(\ref{eqtwo}) is shown in Fig.~\ref{fig14}. The zero Lyapunov exponent of the
response system $x_2(t)$ already becomes  negative as soon as the coupling is
introduced and the onset of CPS is indicated by the negative saturation of the
zero Lyapunov exponent at $b_3=0.11$. The positive Lyapunov exponent of the
response system becomes gradually negative in the transition regime ($b_3 \in
(0.12,0.23)$) and it reaches its negative saturation at $b_3=0.23$ at which high
quality CPS exists.  The transition of the positive Lyapunov exponent to
negativity in this rather complex attractor is again a firm indication of some
degree of  correlation in the amplitudes of both  systems even before the onset
of CPS. As noted earlier, this behaviour of negative transition of positive
Lyapunov exponent of response system before CPS  has also been observed in
Refs.~\cite{mgrasp1996,sgchl2005,bhgvo2003}.

\section{Summary and conclusion}

We have identified and characterized the existence of CPS in both the coupled 
piecewise-linear time-delay systems  and in the coupled Mackey-Glass systems
possessing highly non-phase-coherent chaotic attractors.   We have shown that
there is a typical transition from a non-synchronized state to CPS and
subsequently to GS as a function of the coupling strength in both systems. 
Similar results are obtained for different sampling intervals $\Delta t$ and
for various values of delay time $\tau$.

We have introduced a suitable transformation, which works equally well for both
the systems possessing characteristically distinct attractors (hyperchaotic
attractor in piecewise linear time-delay system and chaotic attractor in the
Mackey-Glass system), to capture the phase of the underling non-phase-coherent
attractor. Both the phase and the frequency locking criteria are satisfied by the
instantaneous phases calculated from the transformed attractors in both the
piecewise-linear and the Mackey-Glass time-delay systems. The frequency ratio
and its difference as a function of coupling strength clearly shows the onset of
CPS in both cases. We have also characterized the existence of CPS and GS in
terms of recurrence based indices, namely generalized autocorrelation function
$P(t)$, CPR, JPR and SPR  and quantified the different synchronization regimes
in terms of them.  The onset of CPS and GS are also clearly shown by the spectra
of CPR, JPR and SPR. The above transition is also  confirmed by the changes in
the spectrum of Lyapunov exponents.   The recurrence based technique as well as
the new transformation are also appropriate for the analysis of experimental
data and we are now investigating the experimental verification of these
findings in nonlinear electronic circuits and in biological systems. Also the
recurrence based indices are found to be more appropriate for identifying the
existence and analysis of synchronizations, in particular CPS, and their onset
in the case of nonlinear time-delay systems in general, where very often the
attractor is non-phase-coherent and high-dimensional. It is also to be
emphasized that the recurrence based measures are more efficient than other
nonlinear techniques~\cite{nmmcr2007} such as mutual information, predictability,
etc. These measures have high potential for applications and
we are also investigating the possibility of extending these techniques to
complex networks.  

\begin{acknowledgments} The work of D. V. S and M. L has been supported by a
Department of Science and Technology, Government of India sponsored research
project.  The work of M. L is supported by a DST Ramanna Fellowship.
J. K has been supported by his Humboldt-CSIR research award and  NoE BIOSIM (EU)
Contract No. LSHB-CT-2004-005137. 
\end{acknowledgments} 
\appendix 
\section{\label{a1} CPS in chaotic systems: Phase-coherent and
non-phase-coherent attractors} 

Definition of CPS in coupled chaotic systems is derived from the classical
definition of phase synchronization in periodic oscillators. Interacting chaotic
systems are said to be in phase synchronized state when there exists entrainment
between phases of the systems, while their amplitudes may remain chaotic and
uncorrelated. In other words, CPS exists when their respective frequencies and
phases are locked~\cite{sbjk2002,aspmgr2001,gvoasp1997}. To study CPS, one has
to identify a well defined phase variable in both coupled systems.  If the flow
of the chaotic oscillators has a proper rotation around a certain reference
point, the phase can be defined in a straightforward way. In this case the
corresponding attractor is referred to as a phase-coherent attractor in the
literature~\cite{sbjk2002,aspmgr2001,gvoasp1997,mgrasp1997,dvskml2006,gvobh2003}
and the phase  can be introduced  straightforwardly 
as~\cite{sbjk2002,aspmgr2001}
\begin{align}
\phi(t)=\arctan(y(t)/x(t)).
\label{ph_1}
\end{align}

A more general approach to define the phase in chaotic oscillators is the
analytic signal approach~\cite{sbjk2002,aspmgr2001} introduced in~\cite{dg1946}.
 The analytic signal $\chi(t)$ is given by
\begin{align}
\chi(t)=s(t) + i\tilde{s}(t)=A(t)\exp^{i\Phi(t)},
\label{ph_2}
\end{align}
where $\tilde{s}(t)$ denotes the Hilbert transform of the observed scalar time
series $s(t)$
\begin{align}
\tilde{s}(t)=\frac{1}{\pi}P.V.\int_{-\infty}^{\infty}\frac{s(t^\prime)}{t-t^\prime}dt^\prime,
\end{align}
where P.V. stands for the Cauchy principle value of the
integral and this method is especially useful for
experimental applications~\cite{sbjk2002,aspmgr2001}.

\begin{figure}
\centering
\includegraphics[width=1.1\columnwidth]{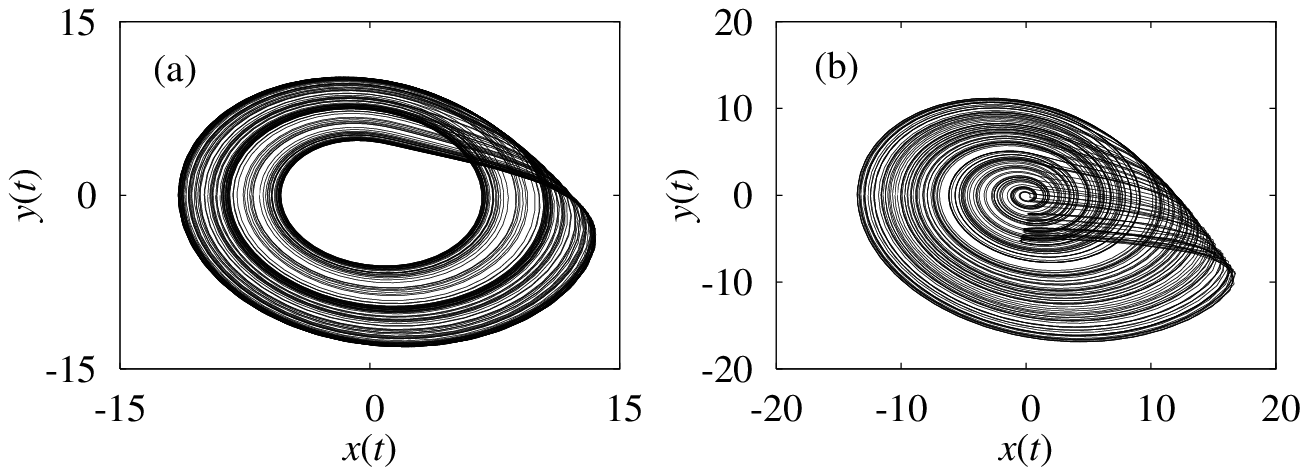}
\caption{\label{fig15} Phase-coherent and funnel (non-phase-coherent) R\"ossler
attractors with parameters (a) $a=0.15$ and (b) $a=0.25$}
\end{figure}

The phase of a chaotic attractor can also be defined based on an
appropriate Poincar\'e section which the chaotic trajectory crosses once for
each rotation.  Each crossing of the orbit with the Poincar\'e section corresponds
to an increment of $2\pi$ of the phase, and the phase in between two crosses is
linearly interpolated~\cite{aspmgr2001,sbjk2002},
\begin{align}
\Phi(t)=2\pi k + 2\pi\frac{t-t_k}{t_{k+1}-t_k},\qquad(t_k<t<t_{k+1})
\label{phase}
\end{align}
where $t_k$ is the time of $k$th crossing of the flow with the  Poincar\'e
section. For the phase coherent chaotic oscillators, that is, for flows which
have a proper rotation around a certain reference point, the phases calculated
by these three different ways are in good agreement~\cite{aspmgr2001,sbjk2002}.

As a typical example, consider the R\"ossler system 
\begin{subequations}
\begin{align}
\dot{x}=&\,-y-z,  \\
\dot{y}=&\,x+ay, \\
\dot{z}=&\,0.2+z(x-8.5).
\end{align}
\label{ros}
\end{subequations}
The topology of the attractor of the R\"ossler system is determined by the
parameter $a$. For $a=0.15$, a phase-coherent attractor (see Fig.~\ref{fig15}a)
is observed with rather simple topological properties~\cite{jdf1980,efs1992},
(where the projection of the chaotic attractor on the ($x,y$) plane looks like a
smeared limit cycle with the phase point always rotates around a fixed origin
with monotonically increasing phase) and hence the phase can be calculated
straightforwardly as discussed above.

However, in chaotic dynamics one often encounters non-phase-coherent
attractors  where the flows are without  a proper rotation around a fixed
reference point (with the origin coinciding with the center of rotation), in
which case a single characteristic time scale does not exist in general.  In
such circumstances it is difficult or impossible to find a proper center of
rotation and it is also intricate to find a Poincar\'e section that is crossed
transversally by all trajectories of the chaotic attractor.    As a consequence
such a non-phase-coherent chaotic attractor is not characterized by a
monotonically increasing phase. Hence phase of such a non-phase-coherent
attractor cannot be defined straightforwardly as in the case of phase-coherent
attractor. Therefore the above definitions of phase are no longer applicable for
non-phase-coherent chaotic attractors.  So  specialized techniques/tools have to
be identified to introduce phase in non-phase-coherent attractors.

It has also  been demonstrated that  certain non-phase-coherent chaotic
attractors can be transformed into smeared limit-cycle like attractors by
introducing a suitable transformation of the original variables.  For example,
in the case of Lorenz attractor, a transformation of the form 
\begin{align}
u(t)=\sqrt{x^2+y^2}
\label{lor}
\end{align}
is introduced~\cite{sbjk2002} and the projected trajectory in the plane ($u,z$)
resembles that of the R\"ossler attractor.  Now phase of the respective
attractor is introduced using the above approaches for phase-coherent
attractors.

However, such a transformation does not always exist or can be found in the case
of non-phase-coherent attractors in general. Again, as a typical example
consider the R\"ossler system specified by Eq.~(\ref{ros}). The topology of the
R\"ossler attractor changes dramatically if the parameter $a$ exceeds $0.21$ and
the phase in this case is not well defined. Funnel (non-phase-coherent)
attractor for the value $a=0.25$ is shown in Fig.~\ref{fig15}b. There are large
and small loops (see Fig.~\ref{fig15}b) on the $(x,y)$ plane and it is not
evident which phase gain should be attributed to these loops and hence phase
cannot be calculated simply as in the case of phase-coherent chaotic 
attractor~(Fig.~\ref{fig15}a) or through simple transformations. Therefore,
recently another definition of the phase based on the general idea of the
curvature has been proposed by Osipov et al~\cite{gvobh2003}. For any
two-dimensional curve ${\bf r} = (u,v)$ the angle velocity at each point is 
\begin{align*}
\nu=(ds/dt)/R,
\end{align*}
where $ds/dt=\sqrt{\dot{u}^2+\dot{v}^2}$ is the speed along the curve and 
$R=(\dot{u}^2+\dot{v}^2)^{3/2}/(\dot{v}\ddot{u}-\ddot{v}\dot{u})$ is the radius
of the curvature.  If $R>0$ at each point, then
\begin{align*}
\nu=\frac{d\Phi}{dt}=\frac{\dot{v}\ddot{u}-\ddot{v}\dot{u}}{\dot{u}^2+\dot{v}^2}
\end{align*}
is always positive and hence the variable
\begin{align}
\Phi=\int \nu dt=\arctan\frac{\dot{v}}{\dot{u}}
\label{curv_phase}
\end{align}
is a monotonically increasing function of time and can be considered as the
phase of the oscillator. These definitions of frequency and phase are general
for any dynamical system if  the projection of the phase trajectory on some
plane is a curve with a positive curvature. Now for the non-phase-coherent
R\"ossler attractor in the funnel regime, the projections of chaotic trajectories on
the plane ($\dot{x},\dot{y}$) always rotate around the origin, and the phase can
be defined  as $\Phi=\arctan(\dot{y}/\dot{x})$~\cite{gvobh2003}.  However, it is
not clear whether an appropriate plane can  always be found, on which the
projected trajectories rotate around the origin for higher dimensional chaotic
systems as such systems will very often exhibit more complicated attractors with
more than one positive Lyapunov exponent as in the case of typical time-delay
systems discussed in the main part of this paper.


\end{document}